\documentclass[useAMS,usenatbib]{mn2e} 
\usepackage{aas_macros}
\usepackage{graphics}
\usepackage[pdftex]{graphicx}
\usepackage{epstopdf}
\usepackage{epsfig}  
\usepackage{natbib} 
\usepackage{dingbat}
\usepackage{float}
\usepackage{amsmath}
\usepackage{times}
\usepackage[varg]{txfonts}                              
\usepackage{verbatim} 
\usepackage{multirow,bigdelim} 
\usepackage{color}
\usepackage{vmargin}
\setmarginsrb{1.2cm}{1cm}{1.2cm}{1cm}{1cm}{1cm}{1cm}{1cm}

\def \msun {M$_{\odot}$}

\usepackage[usenames,dvipsnames,svgnames,table]{xcolor}

  \makeatletter
    \renewcommand{\paragraph}{\@startsection{paragraph}{4}{\z@}%
      {-3.25ex\@plus -1ex \@minus -.2ex}%
      {1.5ex \@plus .2ex}%
      {\normalfont\small\centering}}
     
    \renewcommand{\subparagraph}{\@startsection{subparagraph}{5}{\z@}%
      {-3.25ex\@plus -1ex \@minus -.2ex}%
      {1.5ex \@plus .2ex}%
      {\normalfont\small\centering}}
    \makeatother

\newcommand{\ramses}{{\sc Ramses}}
\newcommand{\clone}{{\sc Clone}}

\newcommand{\music}{{\sc Music}}

\newcommand{\kms}{{ km~s$^{-1}$}}
\newcommand{\hMpc}{{ \textit{h}$^{-1}$~Mpc}}

\title[Virgo formation scenarios]{I- A hydrodynamical \clone\ of the Virgo cluster of galaxies to confirm observationally-driven formation scenarios}
\author[Sorce et al.]
{Jenny G. Sorce$^{1,2}$\thanks{E-mail: \text{jenny.sorce@ens-lyon.fr / jsorce@aip.de}},
  Yohan Dubois$^{3}$, J\'er\'emy Blaizot$^{4}$, Sean L. McGee$^5$, Gustavo Yepes$^{6,7}$,
\and Alexander Knebe$^{6,7,8}$\\
$^1$Univ Lyon, ENS de Lyon, Univ Lyon1, CNRS, Centre de Recherche Astrophysique de Lyon UMR5574, F-69007, Lyon, France\\
$^2$Leibniz-Institut f\"{u}r Astrophysik, An der Sternwarte 16, 14482 Potsdam, Germany\\
$^3$Institut d'Astrophysique de Paris, UMR 7095 CNRS et Universit\'e Pierre et Marie Curie, 98bis Bd Arago, F-75014, Paris, France\\
$^4$Univ Lyon, Univ Lyon1, Ens de Lyon, CNRS, Centre de Recherche Astrophysique de Lyon UMR5574, F-69230, Saint-Genis-Laval, France\\
$^5$University of Birmingham School of Physics and Astronomy, Edgbaston, Birmingham B15 2TT, England\\
$^{6}$Departamento de F\'{\i}sica Te\'orica, Universidad Aut\'onoma de Madrid, Cantoblanco E-28049, Madrid, Spain \\ 
$^{7}$Centro de Investigaci\'on Avanzada en F\'{\i}sica Fundamental,  Facultad de Ciencias, Universidad Aut\'onoma de Madrid, E-28049 Madrid, Spain \\
$^{8}$International Centre for Radio Astronomy Research, University of Western Australia, 35 Stirling Highway, Crawley, Western Australia 6009, Australia\\
}

\begin{document}

\date{}

\pagerange{\pageref{firstpage}--\pageref{lastpage}} \pubyear{2021}

\maketitle

\label{firstpage}

\begin{abstract}
At $\sim$16-17~Mpc from us, the Virgo cluster is a formidable source of information to study cluster formation and galaxy evolution in rich environments. Several observationally-driven formation scenarios arose within the past decade to explain the properties of galaxies that entered the cluster recently and the nature of the last significant merger that the cluster underwent. Confirming these scenarios requires extremely faithful numerical counterparts of the cluster. This paper presents the first \clone, Constrained LOcal and Nesting Environment, simulation of the Virgo cluster within a $\sim$15~Mpc radius sphere. This cosmological hydrodynamical simulation, with feedback from supernovae and active galactic nuclei, with a $\sim$3$\times$10$^7$~M$_\odot$ dark matter particle mass and a minimum cell size of 350~pc in the zoom region, reproduces Virgo within its large scale environment unlike a random cluster simulation. Overall the distribution of the simulated galaxy population matches the observed one including M87. The simulated cluster formation reveals exquisite agreements with observationally-driven scenarios: within the last Gigayear, about 300 small galaxies (M$^*$$>$10$^7$~M$_\odot$) entered the cluster, most of them within the last 500~Myr. The last significant merger event occurred about 2 Gigayears ago: a group with a tenth of the mass of today's cluster entered from the far side as viewed from the Milky Way. This excellent numerical replica of Virgo will permit studying different galaxy type evolution (jellyfish, backsplash, etc.) as well as feedback phenomena in the cluster core via unbiased comparisons between simulated and observed galaxies and hot gas phase profiles to understand this great physics laboratory.

\end{abstract}


\section{Introduction}

\begin{figure*}
\includegraphics[width=0.9\textwidth]{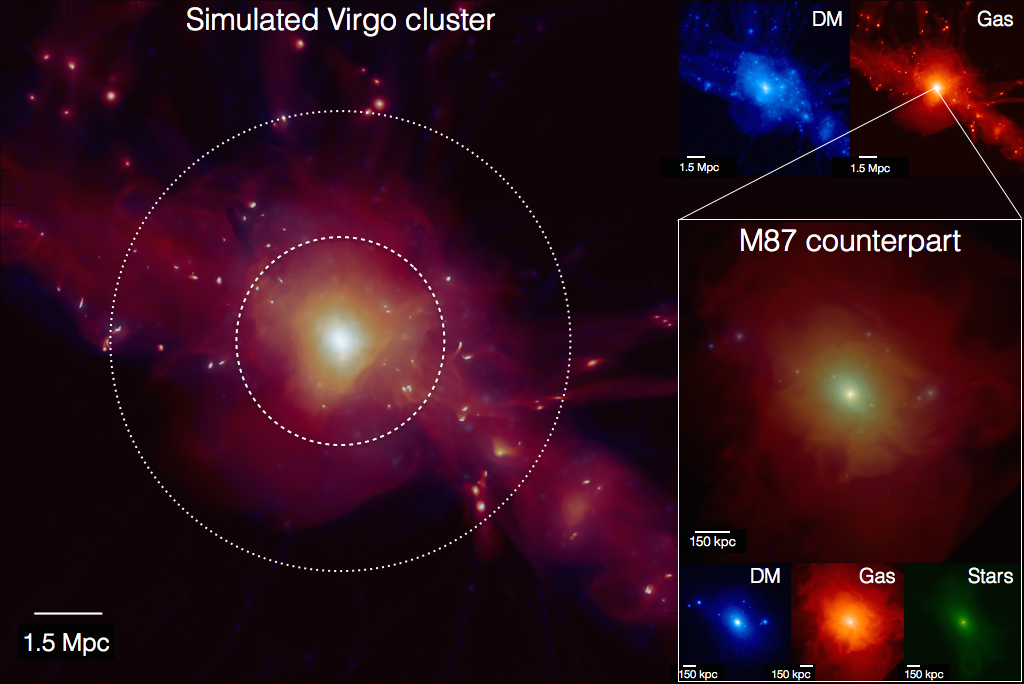}
\caption{Simulated Virgo cluster of galaxies and its central galaxy M87. The largest panel is a combination of dark matter and gas, shown separately in blue and red in the two top right panels. The second largest panel gathers dark matter, gas and stars as seen in blue, red, green in the bottom right three panels. The dashed white line delimits the virial radius while the dotted white line gives the zero velocity radius, i.e. the radius at which the mean radial velocity of galaxies starts to be zero. See the appendix for a more detailed explanation.}
\label{fig:simunice}
\end{figure*}

Since 1936, no less than a thousand titles of publications posted on the Astrophysics Data System (ADS) contain the words 'Virgo cluster', among which more than half are refereed articles. Unconditionally, the Virgo cluster of galaxies, by its proximity with us, has largely been studied both observationally and numerically. The list of papers is long and citing all these papers is unconceivable even if they all deserve some credit. To name a few, we can mention the first paper available in the archive from 1936 
that already gave an estimate of the mass of the cluster \citep{1936ApJ....83...23S}. Quickly the focus turned to dwarf galaxies and their great importance to really understand the cluster's contents \citep{1956AJ.....61...69R}. Then the interest grew towards the (dynamical) distribution of galaxies and the structure of the cluster \citep[e.g.][]{1961AJ.....66..620H,1961ApJS....6..213D} to understand its formation and the evolution of galaxies once they enter this dense environment. Nowadays, the cluster still arouses a lot of interest, explaining the wide diversity seen in a large number of observational projects: e.g. \citet{2000eaa..bookE1822B} for a book review, e.g. \citet{2009eimw.confE..66W,2014ApJ...782....4K} for Spitzer and Hubble studies, \citet{2011MNRAS.416.1996R,2011MNRAS.416.1983R,2016ApJ...823...73L} for multi-wavelength studies, e.g. \citet{2011A&A...528A.107B,2012A&A...543A..33V,2014A&A...570A..69B,2016A&A...585A...2B} for the GUViCS survey, e.g. \citet{2011ASPC..446...77F,2015A&A...573A.129P} for the Herschel Virgo cluster surveys, and e.g. \citet{2012MNRAS.423..787T,2016ApJ...824...10F} for the NGVS survey. We can also mention the black hole of M87, its central galaxy, that became lately famous with the Event Horizon Telescope observations \citep[e.g.][]{2019ApJ...875L...1E,2019ApJ...875L...2E}. There is also a growing number of numerical modeling to understand the cluster \citep[e.g.][for a non-exhaustive list]{1980ApJ...242..861H,2014ApJ...796...10L,2014MNRAS.442.2826C,2014ApJ...792...59Z} but none of these studies to our knowledge both set the cluster within its large scale environment and included baryonic physics. Clearly questions are still arising and a complete picture of the formation and evolution of the Virgo cluster remains to be validated. \\

Indeed, although Virgo is recognized as a dynamically young and unrelaxed cluster with its quite large number of substructures \citep[e.g.][]{1985ESOC...20..181H,1987AJ.....94..251B}, completely understanding the full formation history of the cluster and the evolution of galaxies, once inside this dense environment, from today observations, is largely based on assumptions that require to be tested against modeling. However the diversity of clusters in terms of morphologies, formation history, etc \citep{1988S&T....75...16S} makes the one-to-one comparison observed versus simulated clusters a daunting task. An alternative consists in reproducing directly the Virgo cluster rather than selecting a cluster among a set of random simulated clusters. \citet{1987ApJ...323L.103B} proposed first that option suggesting to constrain the initial conditions of cosmological simulations, namely the Gaussian field, to reproduce rare objects such as massive clusters. The importance of the large scale environment on the formation of the cluster, later implied the necessity of simulating the cluster in its large scale environment. Namely, the necessity of being able to simulate our cosmic neighborhood, the local Universe, arose. \\

Nowadays, several techniques have the ability to reproduce the local large scale structure \citep[e.g.][]{2008MNRAS.389..497K,2013MNRAS.432..894J,2013MNRAS.429L..84K,2014ApJ...794...94W,1990ApJ...364..349D,1991ApJ...380L...5H,1992ApJ...384..448H,1993ApJ...415L...5G,1998ApJ...492..439B,1999ApJ...520..413Z,2008MNRAS.383.1292L,2016MNRAS.457..172L}. However, reaching a precision of a few Megaparsecs to constrain also the clusters, at the linear / non-linear threshold limit is not immediate \citep[see e.g.][]{2018MNRAS.478.5199S}. Recently, \citet{2016MNRAS.460.2015S} proposed new initial conditions constrained with local observational data that lead to dark-matter-only cosmological simulations resembling the local Universe that are in addition valid down to the Virgo cluster scale. Further studies of these dark-matter-only simulations \citep{2016MNRAS.460.2015S,2019MNRAS.486.3951S} proved the robustness of these initial conditions: resulting simulated Virgo clusters present the same type of merging history, including the preferential accretion direction predicted by observations \citep{2000ApJ...543L..27W}, with properties like the center of mass offset with respect to the spherical center, restricted to a given range of values when compared to random simulated clusters in the same mass range. These counterparts of the observed Virgo cluster appear as exceptional tools to check the observationally-driven formation scenarios more thoroughly once baryons are included.\\

Numerous hypotheses regarding the formation of the cluster and the evolution of the galaxies once they entered it indeed remain uncertain. For instance, \citet{2008ApJ...674..742B} estimated the infall within the Virgo cluster to be about 300 galaxies with stellar mass above 10$^7$\msun\ per Gigayear within the last few Gigayears. This infall rate might explain the large amount of galaxies rapidly becoming red \citep{2014A&ARv..22...74B} with galaxies that entered the inner cluster within the last 500 Myr \citep{2016A&A...596A..11B}. Recently, \citet{2018ApJ...865...40L} found indications of a recent group infall into the Virgo cluster, of about 10\% of the mass of the cluster, within the last few Gigayears. This finding is in excellent agreement with the predictions from the dark-matter-only simulations of the Virgo cluster obtained earlier on. Indeed, \citet{2018A&A...614A.102O} concluded their analysis with the last significant merger, about 10\% the cluster mass, occurred within the last few Gigayears. This further strong agreement between our dark-matter-only constrained simulations and the observationally-driven formation scenarios reinforces the argument in favor of simulating a Virgo counterpart with full hydrodynamics and baryons to check that these findings hold with baryonic matter and to push further the comparisons.\\

Indeed, if a clear link between galaxy population and the high density of a cluster environment has been established over the past few years \citep[e.g.][]{2005ApJ...634...51A,2007ApJ...654...53M,hirschmann14}, further investigations revealed that the history of the cluster plays an equal role on the baryonic phase at both the hot gas level and the galaxy population level \citep{2017MNRAS.470..166H}. To fully understand these reservoirs of galaxies or cosmic laboratories to finally use them as cosmological probes, comparisons must be legitimate. There is no room for biases. First and foremost, this first paper of a series confirms in deeper details that the simulated Virgo cluster shares with its observed counterpart similar galaxy distribution at z=0 and formation history.\\

This paper presents the first \clone, Constrained LOcal and Nesting Environment, simulation of the Virgo cluster: a full zoom hydrodynamical simulation of the Virgo cluster counterpart, as shown in Figure \ref{fig:simunice}, obtained with initial conditions constrained only with local galaxy peculiar velocities, namely the Virgo cluster counterpart was not added as an extra density constraint in its forming region. This specificity permits controlling not only the presence of the cluster in its large scale environment at z=0 but also its formation history. In the following, when not specified, Virgo will refer to the observed Virgo cluster. This zoom-in simulation of the Virgo cluster is the first one of its kind and it was produced with the adaptive mesh refinement \ramses\ code \citep{2002A&A...385..337T}. The first section describes the properties of this simulation as well as the building of its initial conditions. The second part of the paper cross-checks the simulated galaxy population distribution against that of the observed Virgo cluster at z=0 and highlights M87 and its simulated counterpart. The last section before concluding confronts observationally-driven formation scenarios and numerical ones. 


\section{The simulation}

Different papers described at length the construction of the constrained initial conditions and the hydrodynamical features used to evolve these initial conditions. This section thus only summarizes the main concepts regarding the constrained aspect to get a Virgo counterpart and the hydrodynamical aspect to get a realistic galaxy population. It also gives the run properties. 

\subsection{Constrained initial conditions} 

In a previous study, we built 200 realizations of the initial conditions of the local Universe that all form a Virgo cluster counterpart \citep{2019MNRAS.486.3951S}. The details of the steps to produce these constrained initial conditions can be found in \citet[e.g.][]{2016MNRAS.460.2015S}. The local observational data used to constrain the initial conditions are 3D peculiar velocities \citep{2014MNRAS.437.3586S} derived with a Wiener filter technique \citep{1995MNRAS.272..885F} from a distance catalog of local galaxies \citep{2013AJ....146...86T} that are mainly grouped \citep[e.g.][]{2017MNRAS.468.1812S,2017MNRAS.469.2859S}, bias minimized \citep{2015MNRAS.450.2644S} and relocated to their progenitors position using the reverse Zel'dovich approximation \citep{2013MNRAS.430..888D} to form the local clusters at z=0 \citep{2018MNRAS.478.5199S}.\\

The detailed study of the 200 simulated Virgos presented in \citet{2019MNRAS.486.3951S} showed that all of our counterparts share a similar historical behavior and properties with a smaller variance than would have a random set of halos within the same mass range. Additionally, they have clearly distinct features from any set of random halos of the same size. For instance, while the random haloes have an average velocity of 463$\pm$207~km~s$^{-1}$, Virgo counterparts have an average velocity of 646$\pm$79~km~s$^{-1}$ barely within the 1$\sigma$ range of the random distribution at z=0 \citep[see Table 1 and 2 of][for other examples and more details]{2019MNRAS.486.3951S}. As a matter of fact, we found that only 30 per cent of the cluster-size random halos comply within 3$\sigma$ with the mean values (radius, velocity, number of substructures, spin, velocity dispersion, concentration, center of mass offset with respect to the spherical center, \emph{all together}) of our Virgo counterparts at z=0, while 97 per cent of simulated Virgos are in their own 3$\sigma$ scatter simultaneously for the properties mentioned before, i.e. a simulated Virgo with one property value in the tail of the property distribution is likely to have all its other property values in the distribution tails. Note that the percentage of random halos in the 3$\sigma$ scatter drops to 18 per cent when adding the requirement of a similar merging history up to z=4, while that of simulated Virgos stays stable. A study based on a more than 20,000 cluster-size random halos revealed that the cluster environment is the main driver for such differences \citep{2020MNRAS.496.5139S}. Hence the importance of using a counterpart, i.e. a cluster in a proper large scale environment, for such studies.\\

Because they are in their proper large scale environment, our 200 simulated Virgos thus share similar properties and histories. Still, among our 200 realizations, we had to select one for resimulation. We chose the most representative of the full sample, i.e. the Virgo cluster counterpart that has the properties, mentioned hereabove, the closest to the average properties of the full sample of 200 simulated Virgos and a merging history in agreement with the mean history of the total sample. There is no need to select on any other aspect, such as a particular feature during the formation history, since, by construction, they share a history with the same features.\\

To avoid periodicity problems in the local Universe-like region, the boxsize of the selected constrained realization is set to $\sim$740~Mpc at z=0 \citep{2016MNRAS.455.2078S}. To decrease the computational cost of the run and since our interest lays solely in the study of the simulated Virgo cluster in this paper, the zoom-in technique, first proposed by \citet{2001ApJS..137....1B} and implemented in \music~\citep{2011MNRAS.415.2101H}, is used with an effective resolution of 8192$^3$ particles for the highest level (here level 13), corresponding to a dark matter particle mass of $m_{\rm DM,hr}\approx$~3$\times$10$^7$~M$_\odot$, within a $\sim$30~Mpc diameter zoom-in region centered on the Virgo cluster counterpart at z=0. Namely, there are more than 35 million dark matter particles within the $\sim$15~Mpc radius zoom-in region at z=0 that are traced back to the initial redshift. This large zoom-in region is motivated by indications from observations that the environment influences some galaxy properties out to several times the virial radius \citep[e.g.][]{2010MNRAS.404.1231V,2012MNRAS.424..232W}. 

\begin{figure*}
\vspace{-1.5cm}
\includegraphics[width=0.9 \textwidth]{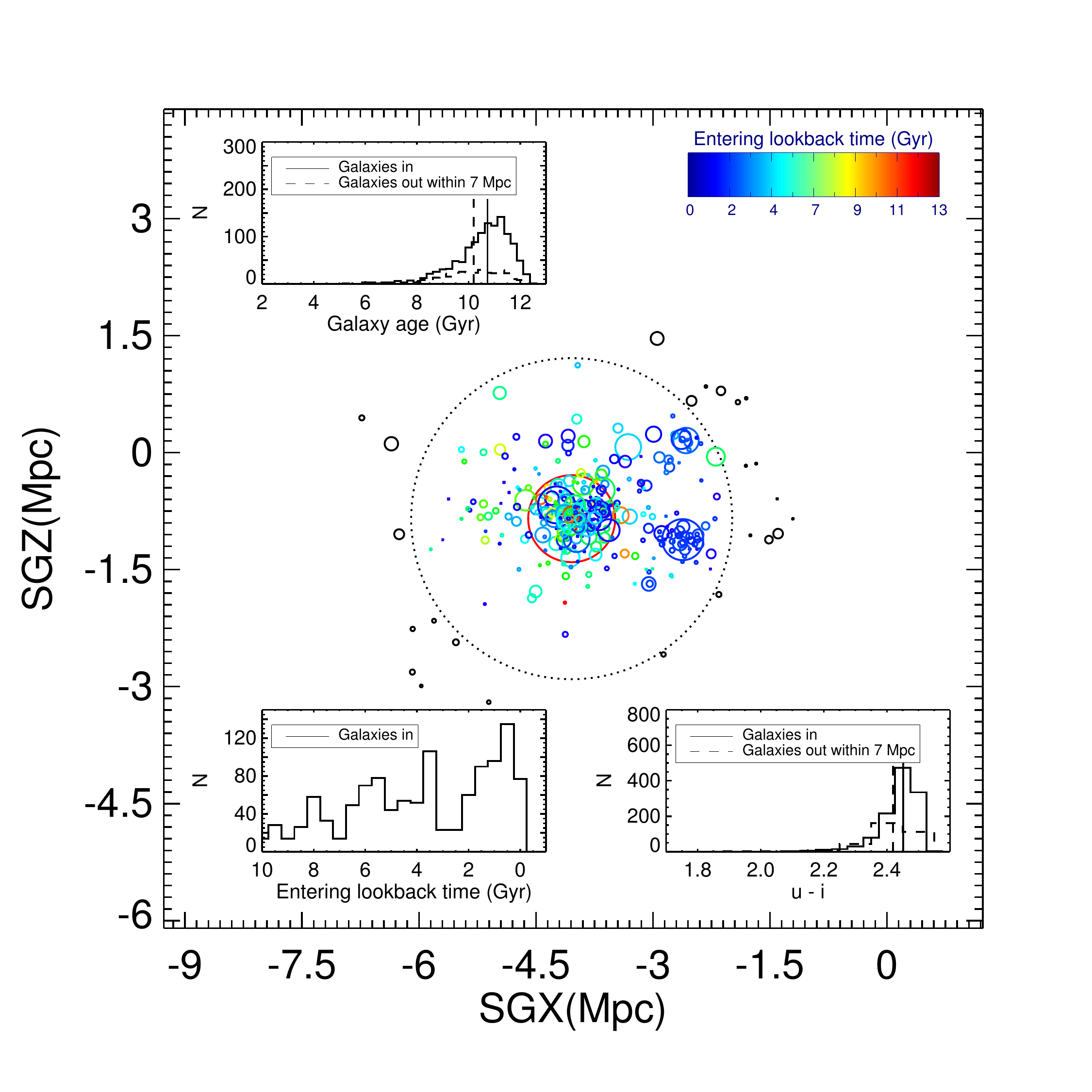}
\vspace{-1cm}
\caption{Main: Simulated galaxies in the Virgo cluster counterpart (i.e. within its $\sim$2~Mpc virial radius) and up to $\sim$3~Mpc from its center, in a tenth of its virial radius ($\sim$0.2~Mpc) thick slice. The dotted circle stands for the virial radius of the simulated dark matter halo at z=0. Colored solid circles represent galaxies. Diameters are proportional to the virial radius of galaxies. Galaxies represented by redder (bluer) circles entered earlier (later) in the cluster counterpart. Galaxies that are not within the virial radius of the cluster at z=0 are in black. Top left and bottom right insets: galaxy age and color distributions inside and outside the virial radius of the cluster as well as their medians. Bottom left inset: galaxy entering time in the virial radius of the cluster. Coordinates are supergalactic coordinates of the Virgo cluster counterpart. Galaxies within the cluster are in general older and redder.}
\label{fig:basic}
\end{figure*}

\subsection{Hydrodynamical features} 

A full set of key physical processes are included as sub-grids models to form a realistic population of galaxies, following the implementation of the Horizon-AGN run \citep{dubois14,2016MNRAS.463.3948D} augmented with black hole (BH) spin-dependent feedback of Active Galactic Nuclei (AGN) with no cluster specific calibration:\\

\noindent
\textbf{Radiative gas cooling and heating} are modeled assuming photo-ionization equilibrium within a homogeneous UV background. The later is imposed from reionization redshift $z_{\rm reion}=10$   following~\cite{1996ApJ...461...20H}. The contribution from metals released by supernovae is included in the cooling curve~\citep{sutherland&dopita93} down to $T=10^4 \, \rm K$. Various synthesized  chemical elements are accounted for but they do not contribute separately to the cooling curve. For simplicity, the tabulated cooling rates rely on the same \emph{relative} abundance of elements as in the sun. The gas follows the equation of state of an ideal monoatomic gas with a 5/3 adiabatic index. \\

\noindent
\textbf{Star formation} is allowed wherever the gas density is greater than 0.1 H cm$^{-3}$ with a random Poisson process spawning stellar particles of mass $m_{\rm s,res}=1.4\times10^{5}\,\rm M_\odot$~\citep{rasera&teyssier06} according to a Schmidt law with a constant star formation efficiency of 0.02 \citep{2007ApJ...654..304K}.\\ 

\noindent
\textbf{Kinetic feedback from type II supernovae} is modeled via a modification of the gas mass, momentum and energy in the surrounding cells \citep{2008A&A...482L..13D} assuming a Salpeter-like initial mass function: release of 10$^{51}$ erg per $10\,\rm M_\odot$ type II supernova with a fraction of the interstellar medium turning into supernovae of $\eta_{\rm SN}=0.2$. Each individual stellar particle deposits $m_{\rm s,res}\eta_{\rm SN}10^{50}\, \rm erg\,M_\odot^{-1}$ at once after 20 Myr with a $0.1$ metal yield with respect to its own content.\\ 

\noindent
\textbf{Active Galactic Nucleus (AGN) feedback} is modeled assuming that a fraction of the rest-mass accreted energy, prescribed by the capped-at-Eddington Bondy-Hoyle-Littleton accretion rate, onto black hole particles, is returned back into the surrounding gas.

A bi-modal\footnote{thermal or kinetic depending on the accretion efficiency.} AGN feedback model is assumed to mimic the quasar-like wind release of i) a radiatively efficient~\cite{shakura&sunyaev73} accretion disc - when the accretion rate is larger than 1 per cent Eddington - or ii) a radiatively inefficient accretion disc, powering collimated jets into their surroundings - when the accretion rate is below 1 per cent Eddington~\citep[see][for technical details of the accretion-ejection scheme]{2012MNRAS.420.2662D}. 

This BH model tracks not only the mass, but also the spin of the BH that can be modified by binary BH coalescence and gas accretion as detailed in~\cite{duboisetal14bhspin1} and~\cite{duboisetal14bhspin2} for the quasar regime and following the magnetically chocked accretion disc solution from~\citet{mckinneyetal12}.

Quasar mode feedback efficiency is set up to $0.15\epsilon_{\rm r}$, with $\epsilon_{\rm r}$ the Eddington rate, in order to reproduce the BH-to-galaxy mass relation~\citep{2012MNRAS.420.2662D}. The efficiency of the jet mode feedback is a function of the spin of the BH according to the solution of~\cite{mckinneyetal12} \citep[see][for more details]{2020arXiv200910578D}.
Note that the radiative efficiency $\epsilon_{\rm r}$ that intervenes in both the Eddington limit and the quasar mode overall efficiency is a function of the BH spin as well.

BH seeds of 10$^5$~M$_\odot$ are formed when the gas density is larger than $0.1\, \rm H\, cm^{-3}$, the local star density is greater than a third of the local gas density and there are no existing black hole within $50 \, \rm ckpc$ \citep{2012MNRAS.420.2662D}.

\subsection{Run properties} 

The simulation ran on 5040 cores using about 10~TB of memory from the starting redshift 120 to redshift 0 during 6 million CPU hours with the adaptive mesh refinement \ramses\ code \citep{2002A&A...385..337T} within the Planck cosmology framework with total matter density $\Omega_{\rm m}$=0.307, dark energy density $\Omega_\Lambda$=0.693, baryonic density $\Omega_{\rm b}$=0.048, Hubble constant H$_0$=67.77\kms~Mpc$^{-1}$, spectral index n$_s$=0.961 and amplitude of the matter power spectrum at 8~\hMpc\ $\sigma_8$~=~0.829~\citep{2014A&A...571A..16P}.

The Euler equations are solved with the MUSCL-Hancock method: a second order Godunov scheme linearly interpolates, with MinMod total variation diminishing scheme, hydrodynamical quantities at cell interface to solve the Euler equations with the approximate Harten-Lax-van Leer-Contact Riemann solver~\citep{Toro1994}.

The initial coarse grid is adaptively refined down to a best-achieved cell size of 0.35~kpc roughly constant in proper length. That is, a new level is added at expansion factors $a~=~0.1,0.2,0.4,0.8$, hence, up to level 21 beyond $a=0.8$. The mesh in the zoomed-in region is allowed to be dynamically refined (or de-refined) from level 13 down to level 21 according to a pseudo-Lagrangian criterion: whenever the total density in a cell is larger (smaller) than $\rho_{\rm DM}+(\Omega_{\rm DM}/\Omega_{\rm b})\rho_{\rm b}>8m_{\rm DM,hr}/\Delta x^3$, where $\rho_{\rm DM}$, $\rho_{\rm b}$ are respectively the DM and baryonic mass densities in the cell of size $\Delta x$, $\Omega_{\rm DM}=\Omega_{\rm m}-\Omega_{\rm b}$ and m$_{\rm DM,hr}$ is the dark matter particle mass for the highest  level. The Particle-in-Cell technique permits deriving the gravitational potential.

A companion simulation with dark matter only using exactly the same features but for the hydrodynamical part was run. Comparisons between this run and the hydrodynamical run are presented in Appendix A. It shows that the hydrodynamical component did not perturb beyond its normal effect the dark matter halo.


 \begin{figure}
 \vspace{-1cm}
\hspace{-0.5cm}\includegraphics[width=0.55 \textwidth]{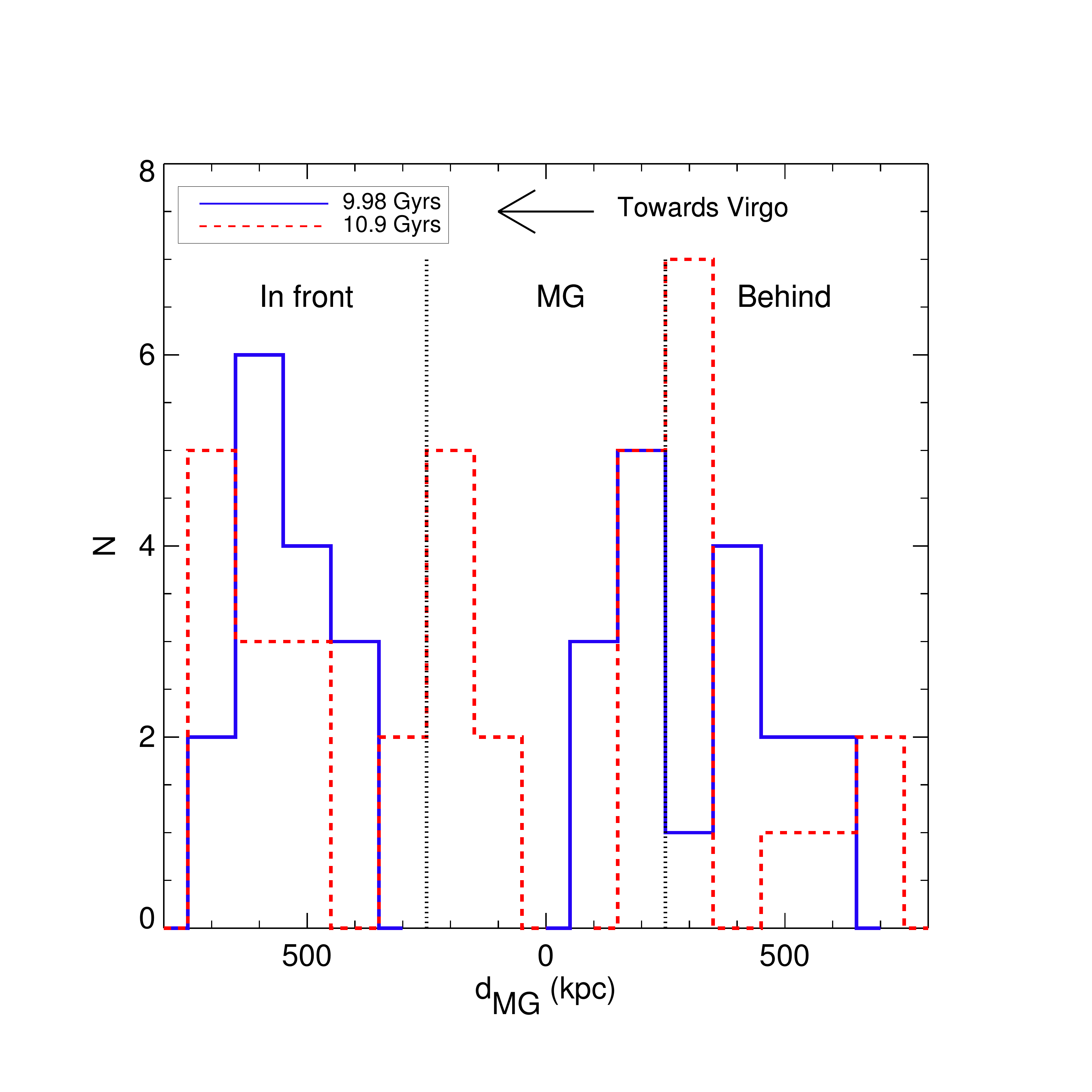}\\

\vspace{-1.2cm}
\caption{Galaxy distributions around the most massive galaxy (MG) of the infalling group when the Universe was about 10 Gyrs (blue solid histogram) and 11 Gyrs (red dashed histogram) old. Galaxies within the Virgo cluster counterpart radius at these given times are excluded. The virial radius of the dark matter halo host of the group main galaxy is about 250~kpc (dotted black lines). The number of galaxies within this main halo (d$_{MG}~<~$r$_{vir}$) increased during that time period. Galaxies are split between those behind this halo with respect to the Virgo cluster counterpart and those between the halo and Virgo counterpart. Clearly, during that time period, galaxies tend to accumulate around MG. The infalling group thus congested during that time period the filament preventing galaxies from falling onto the Virgo cluster counterpart until it finally fell itself.}
\label{fig:channelfil}
\end{figure}

\section{Simulated vs. observed galaxy population}

\begin{figure*}
\hspace{-1.2cm}\includegraphics[width=0.36 \textwidth]{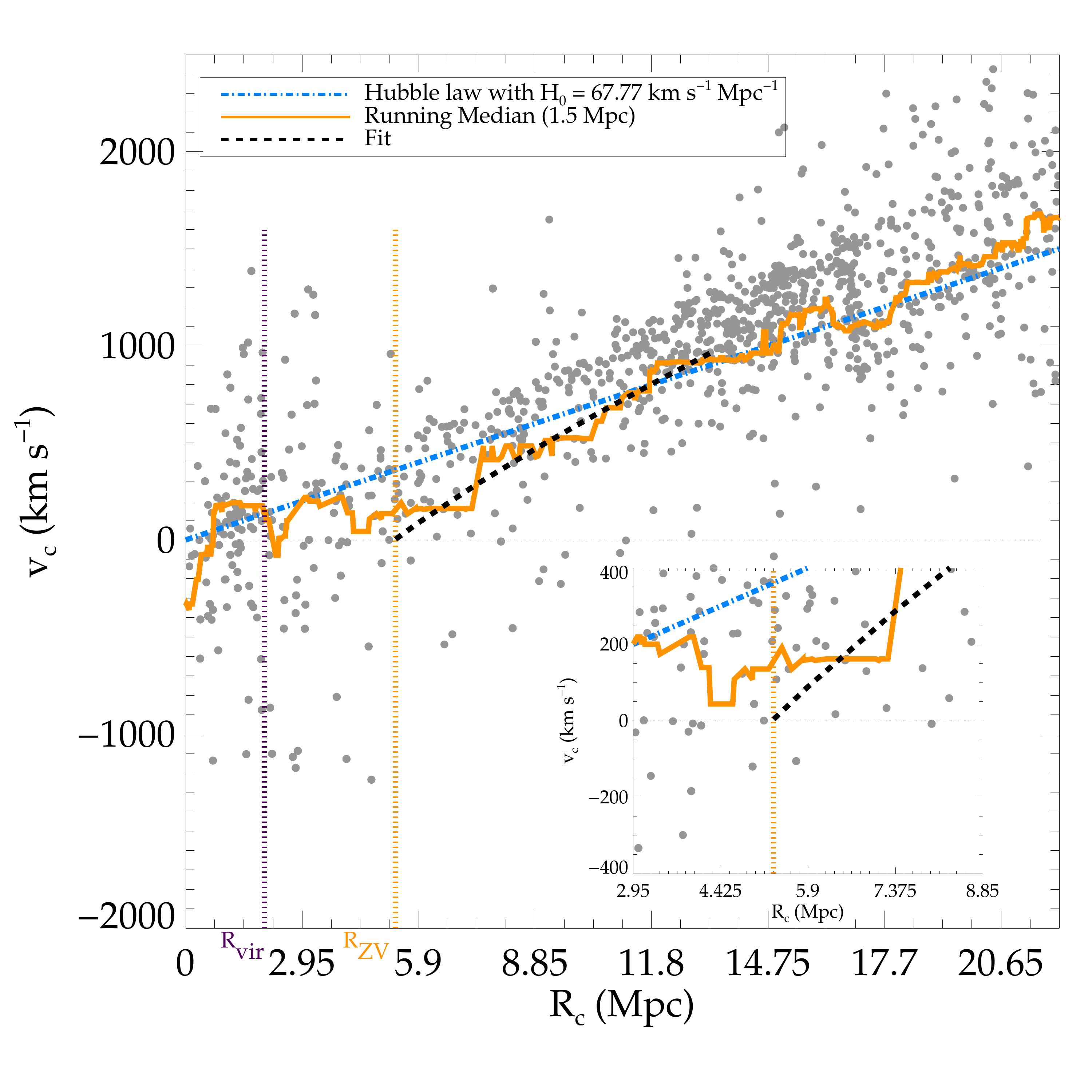}\hspace{-0.18cm}\includegraphics[width=0.36 \textwidth]{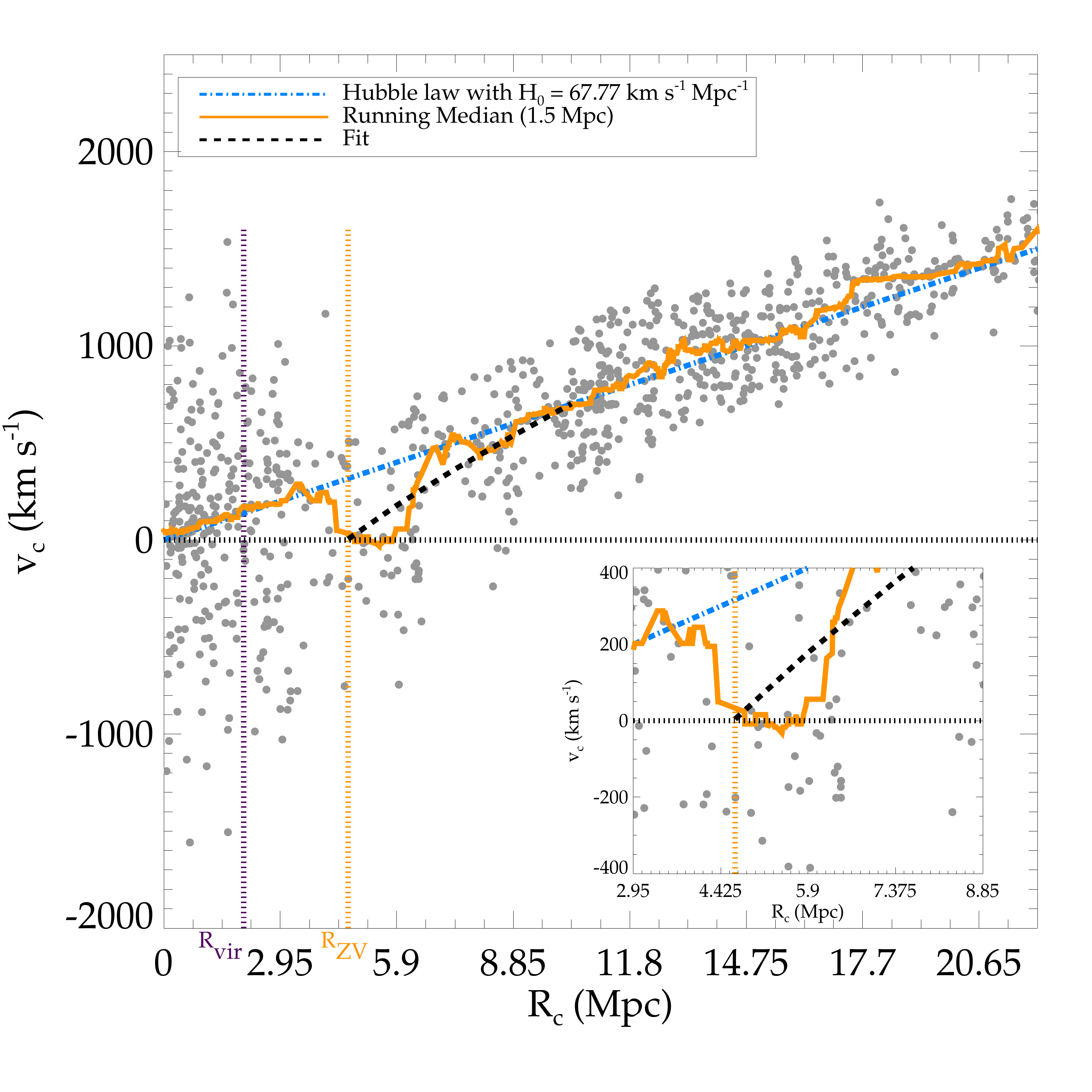}\hspace{-0.3cm}\includegraphics[width=0.36 \textwidth]{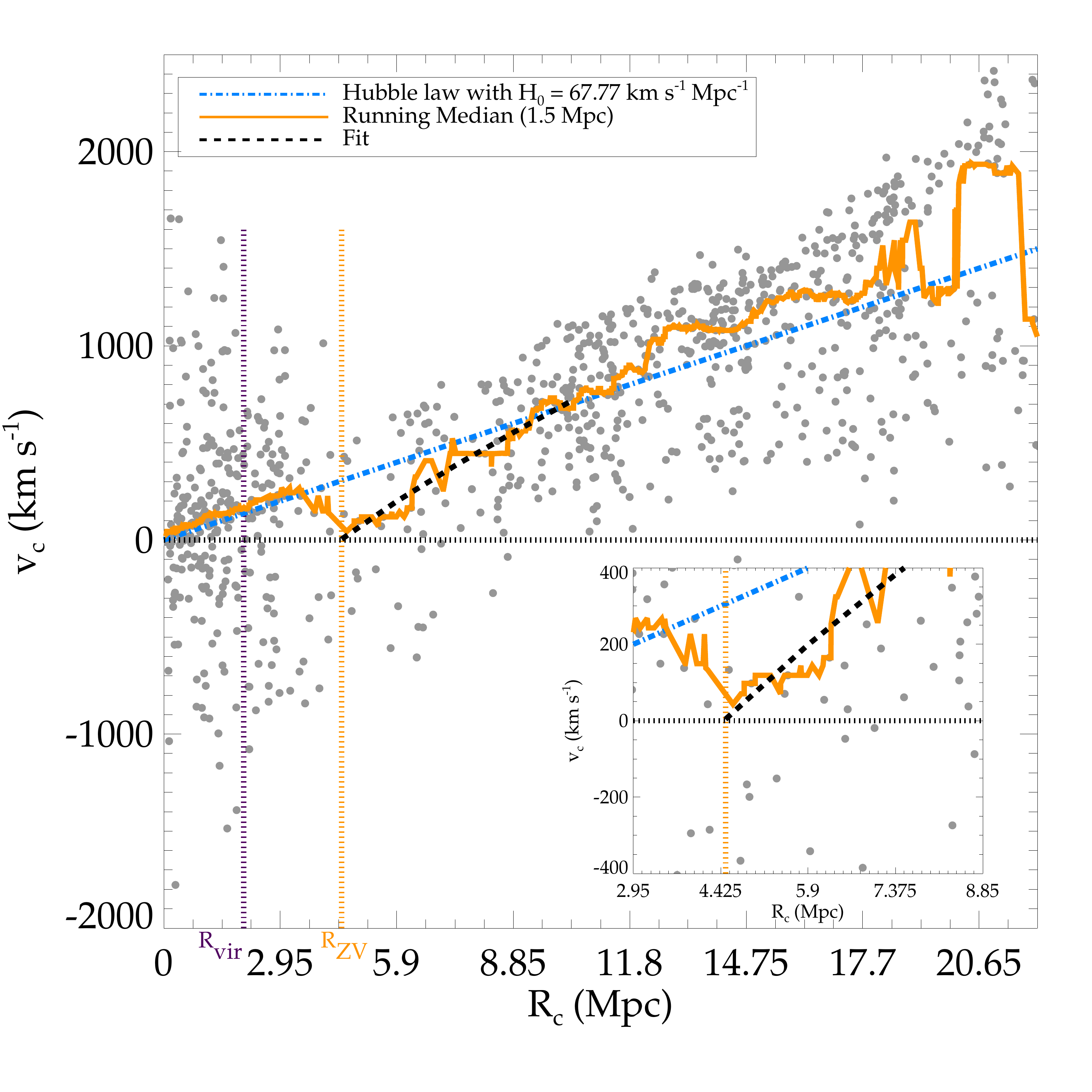}\\
\vspace{-0.5cm}

\caption{Observed data from \citet{2008ApJ...676..184T} (left) and simulated, without (middle) and with (right) error on distances, galaxy velocities with respect to the center of the observed and simulated Virgo cluster. Dotted violet and yellow lines stand for the virial and zero velocity radii. The dot-dashed blue line is the Hubble flow while the dashed black line is a fit to the running median shown with a orange solid line. The small insets are zooms on the zero velocity radius. Only a fraction of the simulated galaxies have been randomly selected to reproduce the same number of galaxies as in the observed sample to properly draw comparisons and estimates of the zero-velocity radius. The dynamical distributions of galaxies in the observed and simulated clusters are extremely similar.}
\label{fig:zerovel}
\end{figure*}

We apply the halo finder algorithm described in \citet{2004MNRAS.352..376A} and \citet{2009A&A...506..647T} to identify galaxies and dark matter halos in the zoom-in region. In this algorithm halos (galaxies) are detected in real space using the local maxima of the density field obtained with dark matter (star) particles. More precisely halos (galaxies) are detected in real space using the local maxima of the density field obtained with dark matter (star) particles. Their edge is defined as the point where the overdensity of dark matter (stellar) mass drops below 80 times the background density. We further apply a lower threshold of a minimum of 100 dark matter (50 star) particles. The particle content of our objects is further `exclusive', i.e. particles belonging to substructures have been removed. That approach allows us to unambiguously investigate the intracluster stellar component. Merger trees permit following galaxies and their evolution across cosmic time, in particular when they cross paths with the cluster. We analyzed 110 snapshots between z=0 and z=7. Then, for each galaxy, we derive the following properties:
\begin{itemize}
\item We define the entering time in the cluster as the time when a galaxy crosses the cluster virial radius at time t for the first time. The later is found via the virial theorem. The later must be verified at a 20\% precision. The specific study of backsplash galaxies at z=0, namely galaxies that entered the cluster recently but exited and did not enter again at z=0, is postponed to a future paper. For information, they represent 5\% of our total galaxy sample, i.e. in a $\sim$12~Mpc radius spherical volume. If they were to be in the cluster virial radius at z=0, the actual number of galaxy within the cluster would be increased by 26\%. 
\item Magnitudes and rest-frame colors of galaxies are derived using single stellar population models from \citet{2003MNRAS.344.1000B} and a Salpeter initial mass function in agreement with the hydrodynamical model we use in the simulation: each star particle contributes to a flux per frequency that depends on its mass, metallicity and age. The contribution of all stars is then summed and filtered to obtain the flux in different bands. Then total luminosities or magnitudes are derived. Since the Virgo cluster is nearby, rest-frame quantities are valid. In this work, attenuation by dust is not included. This implies that observed galaxies have to be corrected for dust extinction before any comparisons with the simulated galaxies. Future work will look in more detail at the dust extinction effect.
\item Three star formation rates are derived:  the amount of stars formed over 10, 100 and 1,000 million years. Depending on the study at in question they will be used alternatively to compare with observations. The specific star formation rate (sSFR) is the ratio of the star formation rate to the current galaxy stellar mass. Note that there is no distinction between stars formed in-situ and those formed ex-situ that have been accreted. The ratio of in-situ / ex-situ stars is a vast topic of research in itself and thus is postponed to future studies.
\item Galaxy metallicity and age are obtained with a sum over all the star particles belonging to the galaxy weighted by their mass. 
\end{itemize}

On a qualitative aspect, Figure \ref{fig:simunice} shows the simulated Virgo cluster as well as its central galaxy M87, today. Their dark matter is in blue while the red color stands for their gas. The stellar density of M87 counterpart is in green. On a more quantitative aspect, Figure \ref{fig:basic} shows a static view of the galaxies in the cluster today in a slice of a tenth of its virial radius. The dashed black line stands for its virial radius and the circles represent galaxies with sizes proportional to their virial radii. The gradient of color from blue to red gives the entering time in the cluster with galaxies that entered earlier in red. The black color stands for galaxies up to 3~Mpc from the cluster center but not within the virial radius of the cluster at z=0. As expected galaxies that entered earlier on are in the cluster core.\\

Three small insets give the galaxy age, entering time and color distributions with medians for galaxies inside and outside of the cluster. As expected galaxies in the  cluster virial radius are on average redder and older than those outside. There is a significant gap in galaxy accretion about 3 Gyrs ago spanning over 1 Gyr. Interestingly it appears just before the beginning of the last significant merger mentioned in the introduction and discussed hereafter. It seems like the infalling group congested during a while the main filament channeling matter into the cluster, hence preventing the accretion onto Virgo counterpart of any other small isolated galaxies for a while. Note that such a preferred direction of infall into the cluster alongside a main filament is observationally confirmed \citep{2000ApJ...543L..27W}. More precisely, Figure \ref{fig:channelfil} shows that the group grew over that period of time, accreting around its most massive galaxy the smaller galaxies that were channeled to the filament. Filaments are indeed known to be the main drivers of matter into the clusters \citep{2019Sci...366...97U}. On this figure, the galaxy distribution around the main galaxy (MG) of the group varies within that time period. Note that galaxies already within the Virgo cluster counterpart at the given time are excluded. First, the number of galaxies within the virial radius of the dark matter halo host of the main galaxy (d$_{MG}~<~$250~kpc) grows (solid blue vs. dashed red line histograms in between the two dotted black lines). Second, always within that time period, galaxies that are behind the group with respect to the Virgo cluster counterpart tend to accumulate at the edge of the dark matter halo virial radius. Third, galaxies between the group and the Virgo cluster counterpart either are accreted onto the simulated cluster or retained by the small group creating a small dip in the galaxy distribution. Hence, fewer galaxies entered within the simulated cluster until the small group itself went in. There is another, although less significant, decrease in galaxy accretion about 7 Gyrs ago. It corresponds to the typical transition for a Virgo-like simulated cluster between an active period of mass accretion, relative to an average random halo, and a quiet history \citep[see][for a detailed explanation]{2020MNRAS.496.5139S}. \\

 \subsection{Simulated vs. observed general distributions}

 \begin{figure*}
 \vspace{-1.2cm}
\includegraphics[width=0.49 \textwidth]{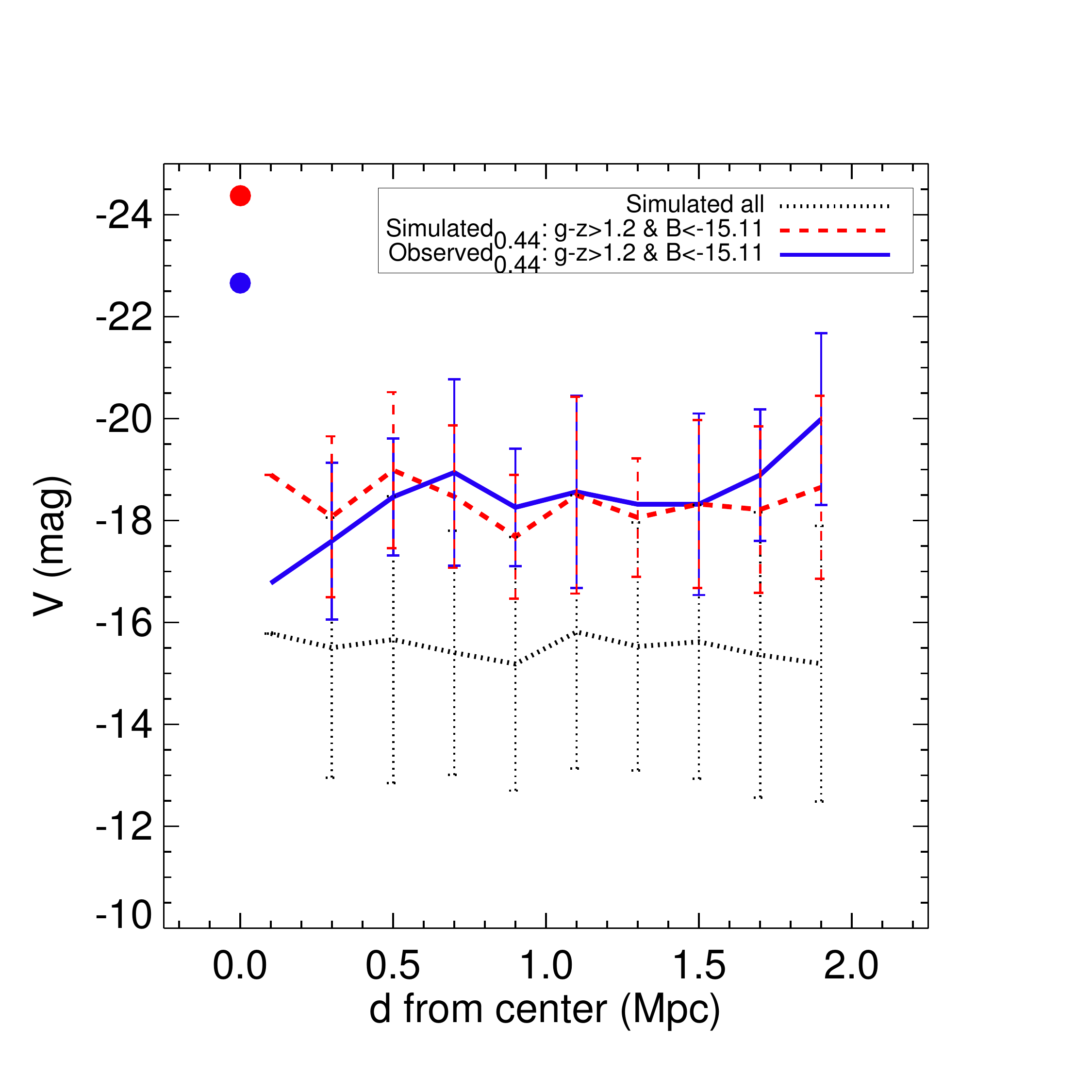}\hspace{-0.5cm}
\includegraphics[width=0.49 \textwidth]{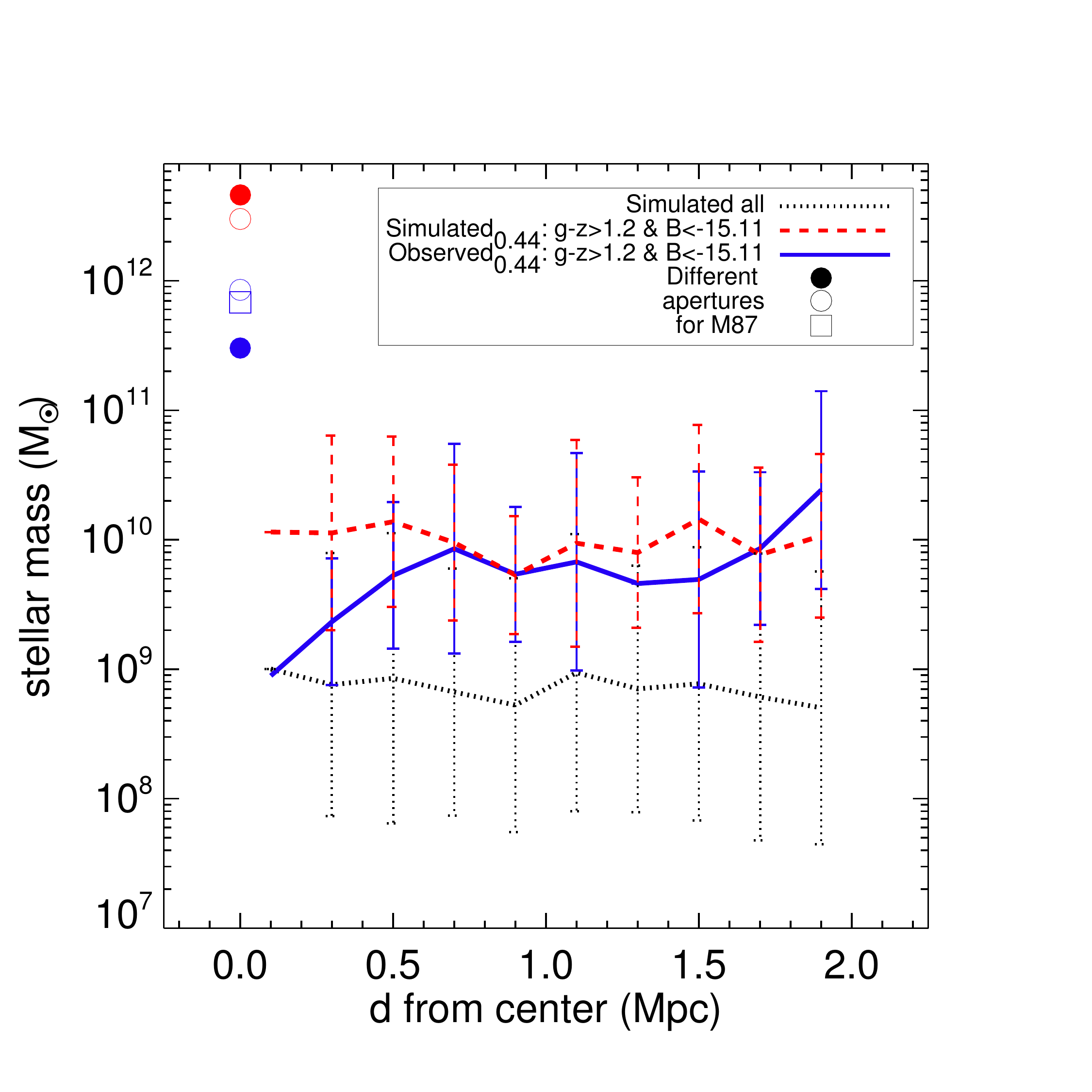}
\vspace{-0.5cm}

\caption{Mean magnitudes (left) and stellar masses (right) of simulated and observed from \citet{2008ApJ...681..197P} galaxies as a function of the distance from the cluster center. The filled circles stand for the observed (blue) and simulated (red) M87. Open symbols stand for stellar mass estimates of the observed (blue) and simulated (red) M87 in different aperture radii as follows:  the open circle is a measure within a $\sim$148~kpc radius from \citet{2009ApJ...700.1690G} and the open square gives the measurements within a $\sim$80.4~kpc radius from \citet{2012MNRAS.421..635F} against a $\sim$5.4~kpc radius for that of \citet{2008ApJ...681..197P}. The dotted black lines represent the full sample of simulated galaxies while the dashed red lines are obtained with the sample restricted according to the same requirements as the observed one. The error bars are standard deviations around the mean. Note that there is no statistics in the observed inner core, most probably because galaxies therein are in the shadow of M87 thus were not selected by the observers. The magnitude and mass distributions of galaxies within the observed and simulated clusters present the same trend. The counterpart of M87 is slightly brighter, thus slightly more massive than the observed M87 by a factor $\sim$3 when compared in the same aperture radius (open circles).}
\label{fig:compobssimu}
\end{figure*}

Velocities and distances in the observed and numerical cluster-centric systems are derived with the formula and geometric considerations given by \citet{2006Ap.....49....3K} (see Appendix B for details). Figure \ref{fig:zerovel} shows the distribution of the velocities as a function of the distance to the cluster center for the observed cluster and for the simulated cluster without and with errors on distances. Because the observed galaxy sample does not fulfill any specific completeness, i.e. the only requirement is to have both a direct distance measurement of the galaxy and its redshift, simulated galaxies are simply randomly selected to match the number of galaxies in the observational sample and facilitate the comparisons with the observed cluster. Additionally, in the third panel of the figure, errors up to 20\% have been added to distances to mimic direct distance measurement uncertainties \citep[e.g.][]{2016AJ....152...50T}. Redshift uncertainties are negligible in this case. Running medians, derived with a window of 1.5~Mpc, are plotted on top of the Hubble diagrams as solid orange lines. The dot-dashed blue line is the Hubble law with  
H$_0$~=~67.77~\kms~Mpc$^{-1}$. The dotted violet and orange lines indicate the virial and zero velocity radii (see Appendix C for the definition of the latter radius). The similarities between the observed and simulated Virgo clusters are remarkable. Actually, the only clear difference between the left and middle panels is in the intrinsic scatter around the Hubble law at large cluster centric distances. This scatter appears larger in the observational case than in the simulation case. It is exclusively due to distance estimate uncertainties in observations that propagate to velocities and that are absent in the simulation as shown by the right panel. \\

\citet{2008ApJ...681..197P} derived the V band magnitude, the distance to the cluster center and the stellar mass of a subsample of galaxies in the Virgo cluster. Stellar masses are obtained combining the (g-z) and (J-K$_s$) colors to simple stellar populations modeling with a Chabrier IMF to get mean mass-to-light ratios in z band. Figure \ref{fig:compobssimu} left shows as a solid blue line the average magnitude by distance bin for this subsample of early galaxies \citep[g-z$>$1.2,][]{2006ApJS..164..334F} with SDSS B band magnitudes smaller than -15.11. In addition, their sample is only complete at the brightest end (-19.1 in SDSS B band). Ultimately, they reach only 44\% completeness in the sample probed here \citep{2004ApJS..153..223C}. The error bars are the standard deviations. The blue circle is M87. The dotted black line stands for the simulated galaxies. To legitimize the comparison, the simulated sample is cut to match the observational restriction: g-z$>$1.2, magnitude larger smaller than -15.11 in the B band and 44\% completeness above -19.1 B band magnitude. Since they claim to have a homogeneous representative subsample, we select randomly 44\% of the galaxies for B band magnitude above -19.1. The dashed red line shows the result. The filled red circle is the counterpart of M87 that happens to be brighter than the observed one. In the latter case, it is important to note however that luminosity asymptotes are not reached before 100-300~kpc for massive central galaxies like M87 \citep{2018AstL...44....8K}. As a consequence, its total magnitude is underestimated. Subsequently, the simulated M87 appears too massive (right panel of the figure). A small aperture indeed leads to an underestimated stellar mass by at least a factor 2 to 4 \citep{2018AstL...44....8K}. In the following a more detailed comparison between M87 and its counterpart investigates more closely the difference. Open blue symbols that give mass estimates in larger aperture radii  \citep{2012MNRAS.421..635F,2009ApJ...700.1690G} already give a hint of  a comparison. The open red circle gives the simulated stellar mass in a similar aperture radius as that of \citet{2009ApJ...700.1690G} (open blue circle).

 \begin{figure}
 \vspace{-1.2cm}
\includegraphics[width=0.49 \textwidth]{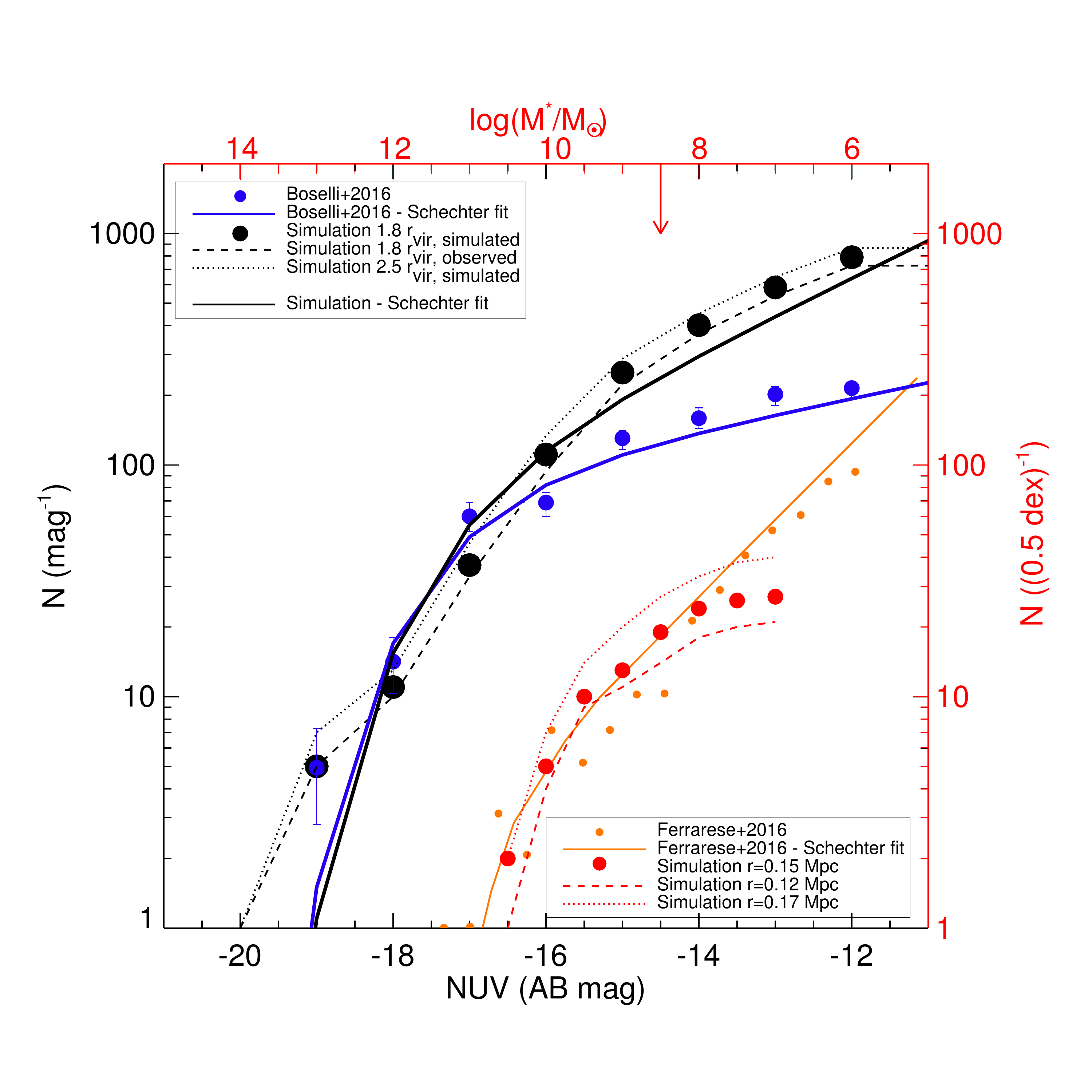}
\vspace{-1cm}

\caption{NUV-luminosity (cold colors) and stellar mass (warm colors) functions of the observed \citep[blue and orange respectively, from][]{2016A&A...585A...2B,2016ApJ...824...10F} and simulated (black and red respectively) Virgo clusters. Different search radii are used in the simulation: filled circles, dashed and dotted lines. Schechter fits are shown as solid lines. The red arrow indicates the stellar mass resolution limit of the simulation. In both cases the simulated function matches relatively well the observational estimates at the brightness / most massive end. The agreement is excellent all the way down to the resolution limit for the stellar mass function.}
\label{fig:nuvlum}
\end{figure}

Coming back to the curves in Figure \ref{fig:compobssimu}, the agreement between observationally and numerically obtained curves and their standard deviations is quite remarkable especially because of the approximate match between the observational and simulated samples. The main difference between the observation and the simulation can be found in the cluster core and at the edges of the sampled volume. Close to the center, there are a few small simulated galaxies while in the observational sample as is, there is no statistics. It could be that the bright M87 makes it challenging to observe small galaxies close to the cluster core. Consequently, in their selection, \citet{2008ApJ...681..197P} did not retain them. Note that the agreement is as impressive between the V band magnitude curves as for the stellar mass curves although they used a Chabrier IMF to derive stellar masses while we used a Salpeter IMF to reversely derive magnitudes. At the observational sample edges, the completeness is most probably affected by distance uncertainties that are about 0.5~Mpc. Still overall the agreement is quite good. \\

To push deeper the comparison, it is necessary to refer to the luminosity and stellar mass functions of the Virgo cluster that are obtained with complete (or corrected for completeness) samples. Figure \ref{fig:nuvlum} reports two of these most recent functions \citep{2016ApJ...824...10F,2016A&A...585A...2B} within different radii: about 0.15~Mpc and 2.88~Mpc (1.8$\times$r$_{vir\, observed}$) . The agreement with their simulated counterparts is remarkable. For the stellar mass function, the observational Schechter fit shown as a solid orange line matches perfectly the simulation (red filled circles) down to the mass resolution limit (red arrow and see Appendix D). As for the luminosity function, identical Schechter functions but for the $\alpha$ slope characterizing the faint end can be used: 
\begin{equation}
\phi(M)=(0.4\mathrm{ln}10)\phi^*[10^{0.4(M^*-M)}]^{(1+\alpha)}\mathrm{exp}[-10^{0.4(M^*-M)}]
\label{eq:sch}
\end{equation}
where M is the absolute magnitude, $\alpha$, $\phi^*$ and M$^*$ are the fitting parameters. $\alpha$ is slightly larger (-1.4 against -1.19) in the simulation than in the observations. Two parameters might come into play: 1) we used a Salpeter IMF to derive retroactively the NUV band magnitude that could give lower (in absolute value) magnitude, shifting galaxies from higher magnitude bins to lower ones, 2) the feedback is not efficient enough at suppressing star formation in the intermediate mass galaxies. An on-going run with a Kroupa IMF based feedback will permit evaluating both of these effects.

In future studies of Virgo counterpart substructures, we will go into more detailed comparisons between their observed and simulated luminosity functions as proposed by \citet{2016A&A...585A...2B}. Before moving to the next subsection, we stress again the excellent agreement between the simulated and observed functions.

  \subsection{Simulated vs. observed M87}

\begin{figure*}
\includegraphics[width=1 \textwidth]{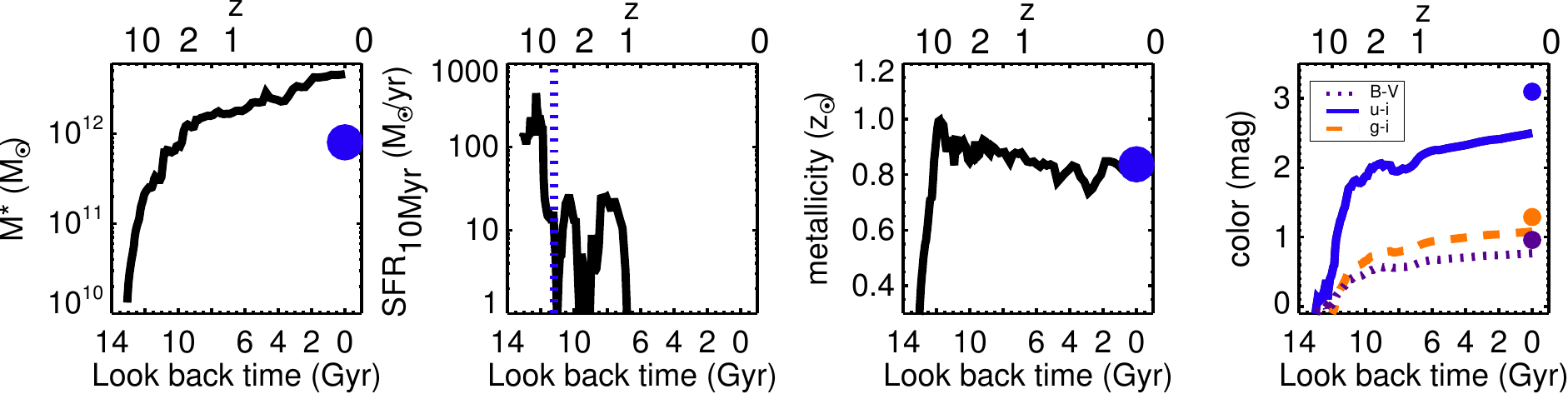}
\caption{Evolution history of the properties of M87 counterpart. Filled blue, orange and violet circles stand for the observational estimates of M87 today: stellar mass from \citet{2009ApJ...700.1690G}, average metallicity and age from \citet{2005AJ....129.2628L}, colors from SDSS \citep{2009yCat.2294....0A} and GUViCS \citep{2007ApJS..173..185G}. The dotted blue line in the second panel stands for the average age of observed stars at the center of M87. Solid lines in each panel as well as the dotted and dashed lines in the last panel are values for the simulated M87 across cosmic time. While the mean age of stars in the counterpart of M87 and its mean metallicity match observational estimates, M87 is redder and less massive than the simulated one.}
\label{fig:m87}
\end{figure*}

 \begin{figure*}
\includegraphics[width=1 \textwidth]{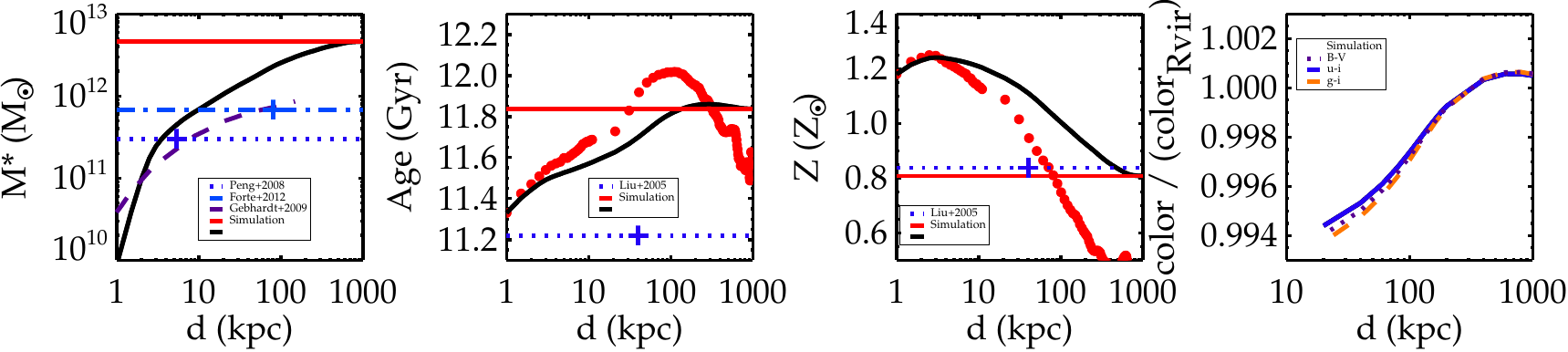}
\caption{From left to right: Solid black lines are stellar mass, age and metallicity cumulative profiles of M87 counterpart. Lines in the last panel stand for the ratio of colors within a given aperture radius (d) to that within the virial radius of the simulated galaxy. Blue lines (but in the last panel) are observational estimates for a given aperture radius marked as a blue cross: stellar mass from \citet{2008ApJ...681..197P,2012MNRAS.421..635F,2009ApJ...700.1690G} with respective aperture radii of 5.4~kpc, 80.4 ~kpc and 148~kpc, average metallicity and age from \citet{2005AJ....129.2628L} using a 40~kpc aperture radius. The violet line in the left panel is the cumulative stellar mass estimate profile of M87. Straight red lines stand for global simulated values for M87 counterpart within its virial radius. Filled red circles give estimates per bin of galactocentric distances. Aperture radii explain part of the discrepancy between the observational and simulated mass and age estimates of M87 but not that between metallicity and color estimates.}
\label{fig:profile}
\end{figure*}

The simulated cluster reproduces overall the spatial and luminosity distribution of observed galaxies within the Virgo cluster.  It is worth pushing further the comparison for M87, its central galaxy.\\

Figure \ref{fig:m87} gives the formation and evolution of M87 counterpart's properties (lines) as well as observational estimates nowadays (filled circles and dotted blue line). Color estimates are from SDSS \citep{2009yCat.2294....0A} and GUViCS \citep{2007ApJS..173..185G}, the stellar mass is from \citet{2009ApJ...700.1690G} and the average metallicity and age are from \citet{2005AJ....129.2628L}:
\begin{itemize}
\item Stellar mass: as mentioned in the previous subsection, bright cluster galaxies usually have stellar masses underestimated by a factor 2 to 4 because their cumulative stellar mass continues growing at the last measured aperture radius. Figure \ref{fig:profile} left shows the cumulative stellar mass profile for M87  \citep{2009ApJ...700.1690G}  and its counterpart as well as stellar masses from \citet{2008ApJ...681..197P,2012MNRAS.421..635F}. Clearly, the latter are lower limits. It confirms that the discrepancy observed between filled red and blue circles in Fig. \ref{fig:compobssimu} is due mostly to the aperture. The estimate from \citet{2009ApJ...700.1690G} is closer to the total stellar mass of M87 but is still on the low side. The asymptotic value is not yet completely reached. This effect combined with residual extinction in the observation and a lack of feedback mitigating star formation in the simulation could explain the residual difference. The black hole in the simulated M87 might start growing too late thus its feedback happens too late. \\
\item Age: the SFR peak (both in-situ and ex-situ stars) of the simulated M87 and the mean age of stars in the observed M87 are a close match. The second panel in Figure \ref{fig:profile} confirms that simulated and observational ages are similar especially since observational estimates are measured within less than 40 kpc. \\
\item Metallicity: observed and simulated metallicities of M87 are really close. However,  Figure \ref{fig:profile} shows that, because of the same small aperture radius (40 kpc) used to measure the metallicity, there is in fact a discrepancy and they cannot be considered identical. In a third paper, we will show that the simulated hot gas has a too low metallicity with respect to the observed one, an explanation can be partly found here: metals are insufficiently released by stars.\\
\item Colors: the discrepancy is the most import for the (u-i) color. However, the SDSS u band might not be the best at measuring luminosity of BCGs. Indeed it has a higher systematic uncertainty \citep{2008ApJ...674.1217P}, it is more susceptible to dust extinction, and it is nearer to the unexplained UV upturn which appears in elliptical galaxies \citep{2011MNRAS.410.2679L}. The discrepancy is somehow of a lesser extent for other colors such as (g-i) and (B-V). In these cases, a small residual extinction in observations might explain part of the difference. A lack of feedback dampening star formation in the simulation could also perhaps explain the bluer color for the simulated M87 than the observed one. However, removing the youngest stars in the simulated galaxy does not drastically change its color (by +0.01 at most with the minimum age at 10~Gyr). In addition, the simulated color profile shown in Figure \ref{fig:profile} shows that there is no drastic change with the aperture radius either and in any case it makes the simulated galaxy even slightly bluer (by about -0.01).
\end{itemize}

Besides M87 associated to the cluster A \citep{2014A&A...570A..69B}, the Virgo cluster hosts other old massive galaxies like M60 (cluster C) and M49 (cluster B) with respective mass estimates of roughly about 10$^{12}$M$_\odot$ and 5$\times$10$^{11}$M$_\odot$ at distances of 0.4 and 0.6~Mpc from M87 with (g-i) color of 1.34 and 1.30 mag (values obtained with SIMBAD queries at CDS). 
A quick look in the simulation reveals two old galaxies (11.1 and 11.3 Gyrs) of masses 6$\times$10$^{11}$M$_\odot$ and 3.5$\times$10$^{11}$M$_\odot$ at 0.3 and 0.4~Mpc from M87 counterpart with (g-i) color of 1.06 and 1.05 mag. Without being close matches to M60 and M49, their presence is interesting and further studies with their associated large substructures  \citep{2014A&A...570A..69B} within the simulated Virgo cluster will be the subject of a future paper. Note that the shift in colors is found again and seems systematic. Simulated old massive galaxies appear definitively bluer than their counterparts, comforting the too low AGN feedback hypothesis.


\section{Formation history: from observationally-driven to numerically-confirmed formation scenarios}

\subsection{Accretion rate of galaxies}
 
 \begin{figure}
 \vspace{-1cm}
\hspace{-0.5cm}\includegraphics[width=0.5 \textwidth]{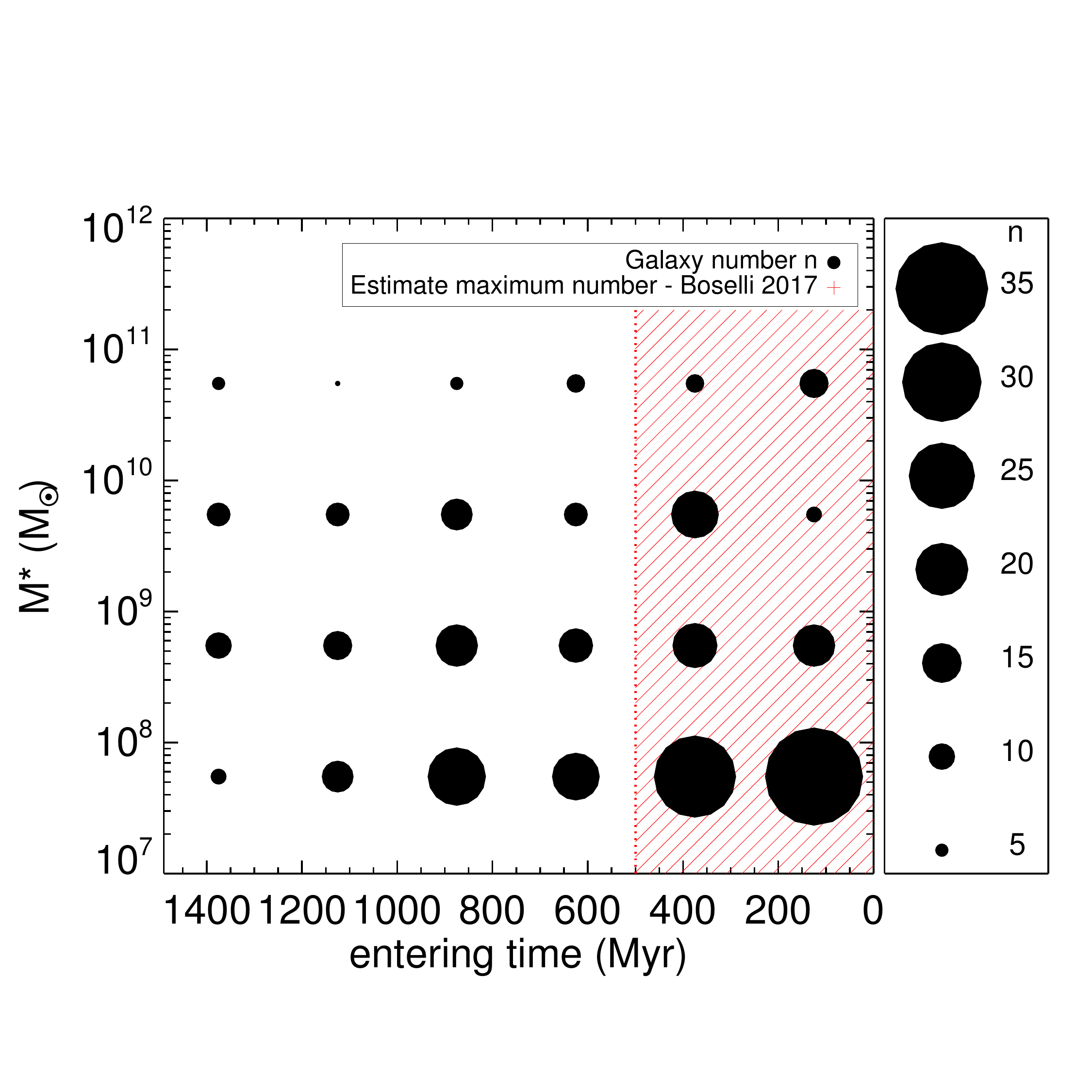}
\vspace{-1cm}

\caption{Number of galaxies per today mass bin and per 250 Myr that entered the cluster. Sizes of filled black circles are proportional to that number. The dashed area stands for the period of time within the last few Gyrs when there should be a maximum number of small galaxies entering the cluster according to observations \citep{2008ApJ...674..742B}. About 300 small galaxies entered the simulated cluster lately mostly within the last 500~Myr in agreement with observationally-driven scenarios.}
\label{fig:entering}
\end{figure}

The overall good agreement between observations of the Virgo cluster and the simulated one allows us to further compare the observationally-derived formation scenarios and the numerical findings. 

 \citet{2008ApJ...674..742B} deduced from observations that about 300 galaxies with stellar mass above 10$^7$\msun\ should have entered the cluster each Gigayear within the last few Gigayears. From observing that a large amount of small galaxies are quickly turning red in the inner cluster, \citet{2014A&ARv..22...74B} even concluded that most of them should have entered within the last 500 Myr  \citep{2016A&A...596A..11B}. According to Figure \ref{fig:entering},  279 galaxies with stellar masses between 10$^7$ and 10$^{11}$ M$_\odot$ entered the cluster within the last Gigayear. Correcting for completeness (see Appendix D), it means that between 280-330 simulated galaxies with masses ranging from 10$^7$ to 10$^{11}$ M$_\odot$ entered the cluster within the past Gigayear (depending on when the galaxies -- missing within the virial radius of the simulated cluster because of the lack of completeness at these masses -- entered the cluster). It is also clear that lately, the simulated Virgo cluster did not accrete major galaxy groups but mostly isolated small galaxies as already shown in our previous studies but based on dark-matter-only simulations \citep[e.g.][]{2018A&A...614A.102O}. Additionally, the figure confirms that a lot of these small galaxies entered the cluster recently with a maximum within the last 500~Myr just like observationally-deduced.\\

To summarize, this paper confirms that results obtained within a series of paper based on dark-matter-only simulations of the Virgo cluster are not affected by the inclusion of baryons and it also goes further. In agreement with observations:
\begin{itemize}
\item the simulated Virgo cluster has had a quiet merging history within the last few Gigayears while it used to grow faster \citep[see][for the DM study]{2019MNRAS.486.3951S},
\item about 300 small isolated galaxies entered recently in the cluster, most of them within the last 500~Myr. 
\item Virgo counterpart accretes along a preferential direction: the line-of-sight filament \citep[see][for the DM study]{2016MNRAS.460.2015S},
\item it underwent its last significant merger within the last 4 Gyrs,
\item this merging event finished a bit more than a Gigayear ago , 
\item it was about 10\% the mass of the cluster today \citep[for these last three points see][for the DM study]{2018A&A...614A.102O}.
\end{itemize}

It is interesting to further look at this last merging event.

\subsection{The last significant merger}

Recently, \citet{2018ApJ...865...40L} found evidence in their observations of the Virgo cluster that, indeed, a group of about 10\% the mass of the cluster probably fell onto it within the last 2 to 3 Gyrs. If it did fall, this group did so along the line-of-sight, namely it entered the cluster directly on the opposite side from ours.  Such a scenario would explain the clustering of quenched galaxies, with high velocities with respect to the cluster center velocity, that are not within the inner core of the cluster in a phase-space diagram. Figure \ref{fig:basic} shows the supergalactic XZ plane that permits making visible the simulated last significant merger of about 10\% the mass of the cluster today at about (-3,-1.5)~Mpc and Figure \ref{fig:channelfil} shows the gathering of this group. This group happens to have fallen into the cluster through the filament opposite to the center of the box with respect to the simulated Virgo cluster, namely quasi in the line-of-sight of an observer close to the center of the box (where the Milky Way would be), in excellent agreement with the observationally-deduced merging history of this group.\\

\begin{figure*}
\vspace{-1cm}
\hspace{-0.5cm}\includegraphics[width=0.35 \textwidth]{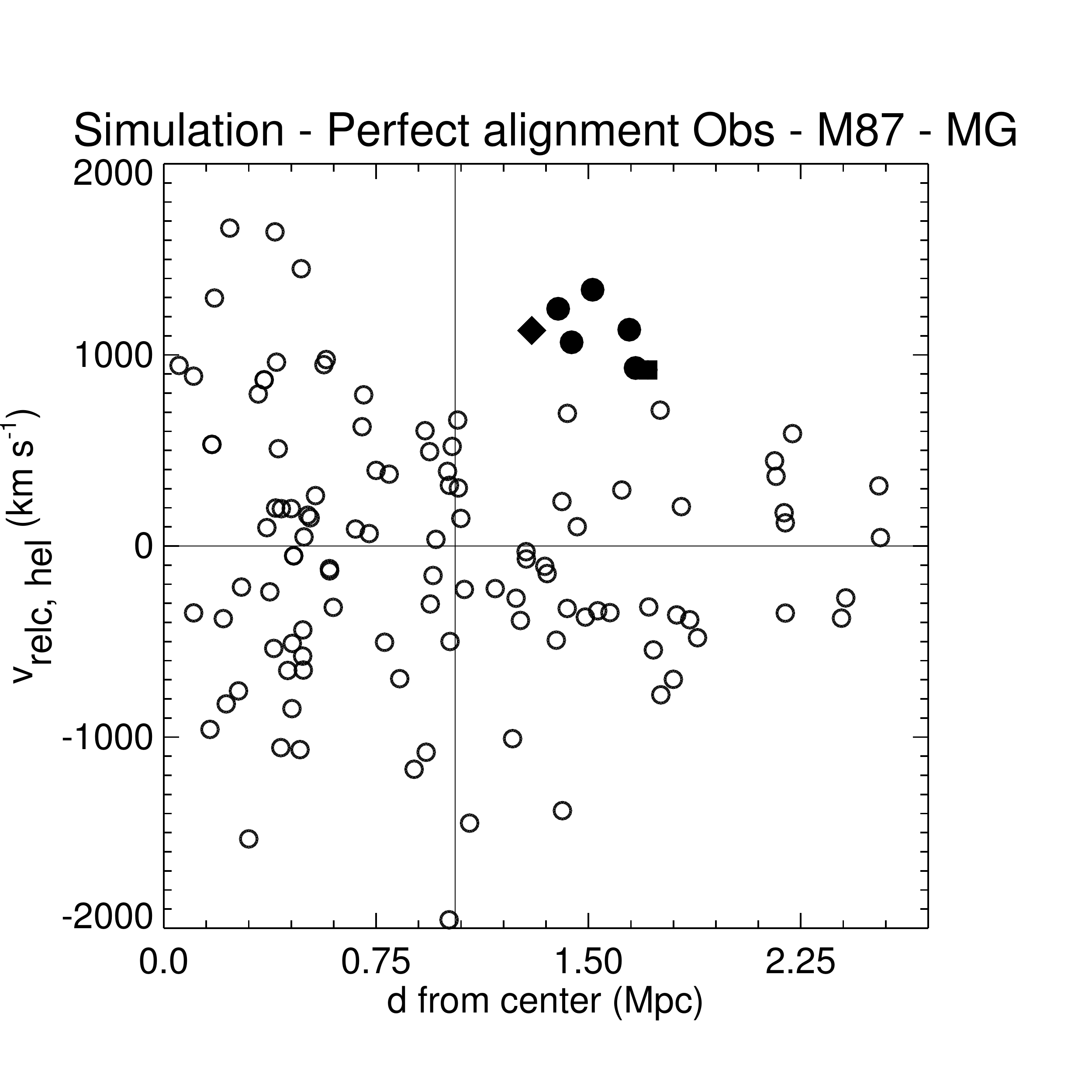}\hspace{-0.4cm}
\includegraphics[width=0.35 \textwidth]{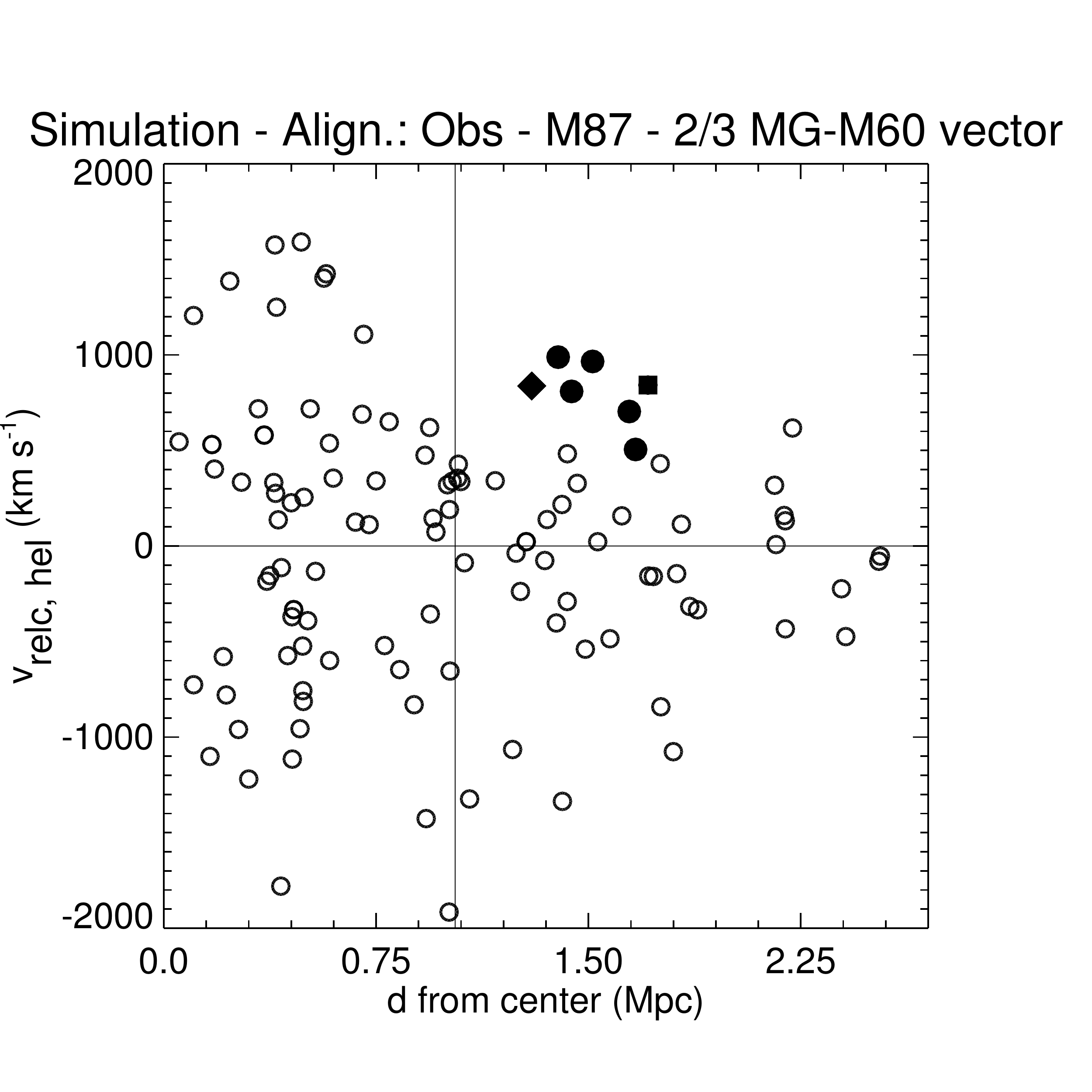}\hspace{-0.4cm}
\includegraphics[width=0.35 \textwidth]{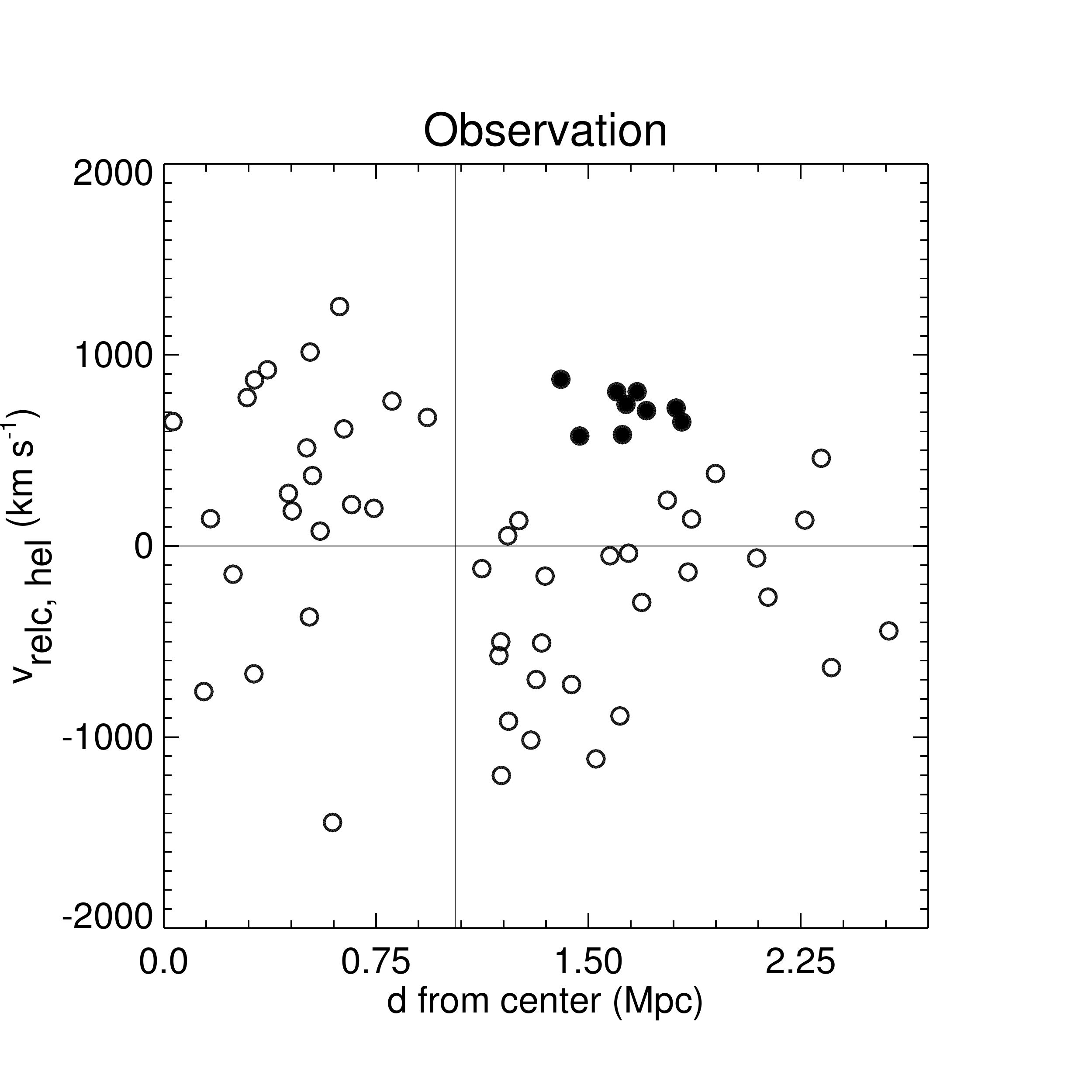}
\vspace{-0.7cm}

\caption{Galaxy velocity with respect to the cluster center velocity as a function of the distance to the center for the simulated (left and middle) and observed \citep[right, from][]{2018ApJ...865...40L} clusters derived in the heliocentric frame of reference.  Left: Circles are simulated galaxies with (g-z)~$>$~1.2. Fully filled black circles denote galaxies that belonged to the group of about 10\% the mass of the cluster that merged within the last few Gigayears. The main galaxy (MG) of the group is taken to be exactly aligned with respect to M87 counterpart and the observer to derive relative velocities. Middle: same as left panel but the observer is taken to be aligned with M87 and 2/3 of the MG-M60 vector.  In these two left panels, the filled diamond stands for a galaxy that fell into the cluster after the group, the filled square represents a galaxy that fell into the cluster before the group. Right: datapoints from the right panel of Fig. 4 in Lisker et al. 2018. Circles are early type galaxies in the observed Virgo cluster. Filled black circles are potential galaxy candidates that used to belong to a group that merged within the cluster a few Gigayears ago. The observational scenario of the merged group somewhat quasi along the line-of-sight seems confirmed.}
\label{fig:losinfall}
\end{figure*}

More precisely, Figure \ref{fig:losinfall} left reproduces with the simulated galaxies the study conducted by \citet{2018ApJ...865...40L} with observational data. Positioning an observer directly in the line-of-sight of the fallen-in group with respect to the cluster, radial velocity in the heliocentric frame of reference (see equation \ref{eq:1} for a conversion from the CMB frame of reference) relative to the mean velocity of the cluster in the heliocentric frame of reference as a function of the distance to the center are derived for the galaxies. Circles represent red galaxies (g-z$>$1.2). Galaxies are selected in a r band magnitude range [-19,-18]. Fully filled circles indicate galaxies that belong to the group of $\sim$10\% the cluster mass that fell into the cluster within the last few Gigayears in perfect alignment with the cluster and the observer.
 
\noindent Conversion from the CMB frame of the reference to the heliocentric one:
\begin{equation}
\vspace{-0.15cm}
v_{cmb}=v_{helio}+v_{\rm apex}({\rm sin}(b){\rm sin}(b_{\rm apex})+{\rm cos}(b){\rm cos}(b_{\rm apex}){\rm cos}(l-l_{\rm apex}))
\label{eq:1}
\end{equation}

where $v_{\rm apex}$=371~km~s$^{-1}$, $b_{\rm apex}$=48.26~deg and $l_{\rm apex}$=264.14~deg from \citet{1996ApJ...473..576F}, l and b are galactic coordinates.

Comparisons with the phase-space diagram in Fig. 4, second row, of \citet{2018ApJ...865...40L} and reproduced in the right panel of Figure \ref{fig:losinfall} lead to exceptional similarities. The group members are found at about the same position as the proposed fallen-in galaxy group in the observational data. Note that the magnitude bin is lower by one unit in the simulated case because of the opposite trend in the red than in the blue case of the Salpeter IMF choice to derive reversely r band magnitudes from stellar masses. In our case, it gives higher (in absolute value) magnitudes for the same given stellar mass.\\

One may notice that relative velocities obtained by \citet{2018ApJ...865...40L} are somewhat smaller than simulation-based relative velocities. Tests changing the line-of-sight in the simulation show that this shift is probably due to the non-perfect observer -- M87 -- MG alignment in the observation. For instance, a simulated alignment observer -- M87 -- 2/3 of the MG-M60 vector as shown in the middle panel permits recovering velocities of the same order as that of   \citet{2018ApJ...865...40L}.

In both cases (perfect and non exact alignments), two simulated galaxies appear close to those belonging to the group in the phase space diagram (filled diamond and square). The galaxy represented by a filled square fell just before the group. The other galaxy, shown as a filled diamond, followed the group's footsteps but in complete isolation from the latter. Interestingly, it appears to be a backsplash galaxy. This is not at odds with observations since galaxies within this region in the phase space diagram are only plausible, but not guaranteed, members of the fallen-in group. Note that this latter galaxy is one of the galaxies that are represented at 700~kpc behind the main galaxy of the group in the histogram of Fig. \ref{fig:channelfil}. \\
 
\begin{figure*}
\vspace{-3.cm}
\hspace{-2cm}\includegraphics[width=1.15 \textwidth]{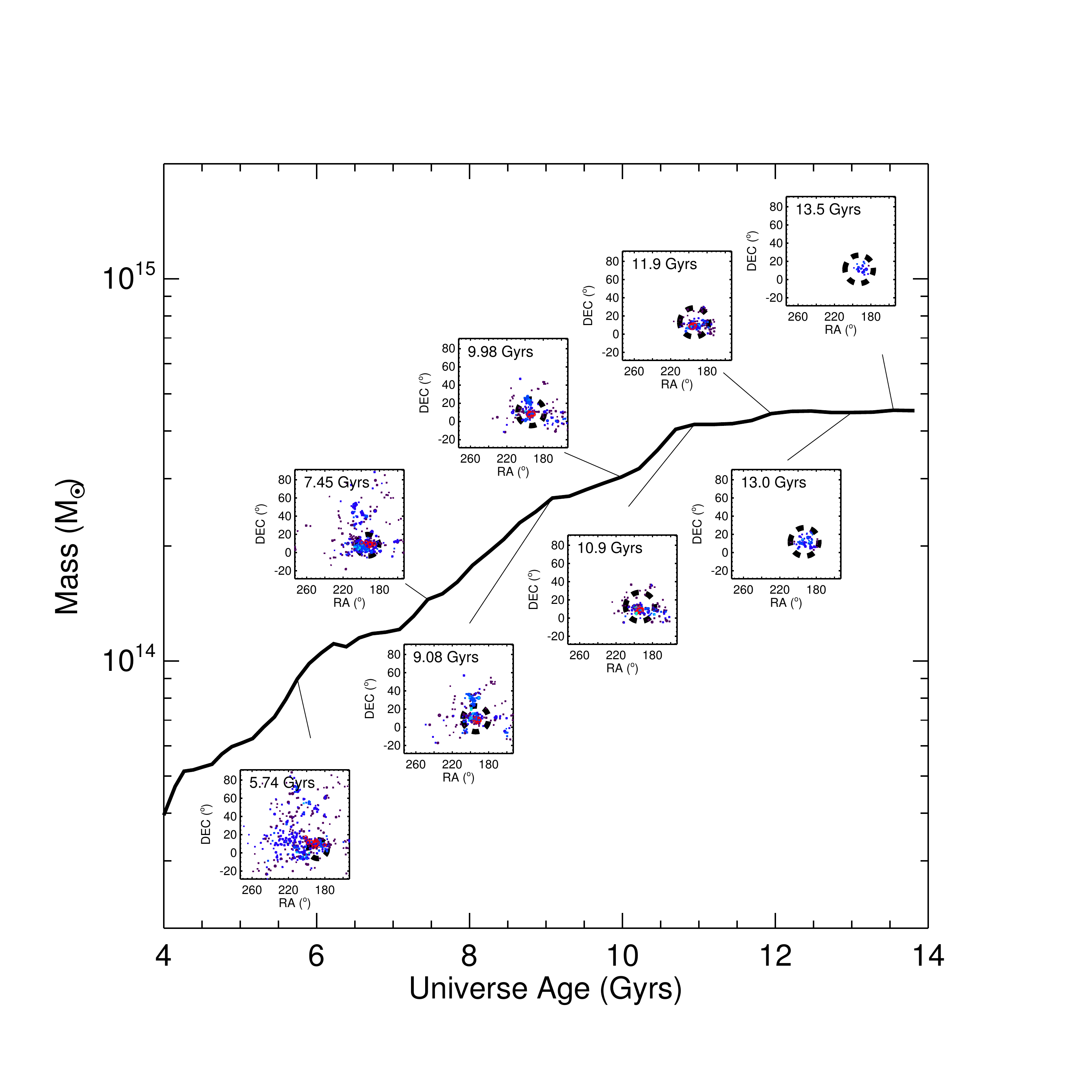}
\vspace{-1.7cm}
\caption{Merging history of the simulated Virgo cluster. Small insets show the position of the galaxies that will end up inside the virial radius of the cluster at redshift zero. The gradient of color gives the status of the galaxy before entering the cluster: violet means that the galaxy was isolated, light blue stands for a satellite galaxy within a group. The red color is used to highlight galaxies from the last significant merging group. The dashed black lines stand for the virial radius of the dark matter halo of the cluster at different epochs in angular coordinates. The factor derived at z=0 to estimate the angular diameter given the distance of the cluster today is also used for earlier redshifts, in other words the simulated cluster is always seen by the fictive observer from the same distance that at z=0. Galaxies within the cluster virial radius at a given time are not shown. The reconstruction of the Virgo cluster formation from observations is matched with good accuracy by the simulation: last significant merger quasi along the line-of-sight, quiet accretion lately and more active earlier on.}
\label{fig:accre}
\end{figure*}

Finally, Figure \ref{fig:accre} summarizes the accretion history of the Virgo cluster as would be seen by an observer at the center of the box (where the Milky Way would be). Similarly to what we observe today, lately the accretion happens to be along a preferential direction of infall (filament in the line-of-sight: galaxies that fell in lately appear to be mostly localized behind the Virgo cluster counterpart represented by a dashed line), quieter than it used to be, with the last significant merger of about 10\% the mass of the cluster that started about 2 Gigayears ago while very recently only about 300 galaxies per Gigayear entered the cluster. Each small inset shows the galaxies that are in the cluster today at an anterior time, namely before they enter the cluster. They are not displayed after they entered the cluster. The thick dashed black line stands for the virial radius at a given time in angular coordinates. The factor used to derive the angular diameter of the cluster given its distance today is also used for earlier redshifts, in other words the angular size is always computed from the same distance that at z=0. The violet color stands for isolated galaxies before entering the cluster while the light blue color means that a galaxy is a satellite within a group before entering the cluster. The red color highlights the galaxies that belong to the last significant merging group.
 

\section{Conclusion}

The Virgo cluster of galaxies is our closest cluster-neighbor. As such, it is a formidable object of study to understand cluster formation and the evolution of galaxies once they enter this rich environment. To that end simulations of galaxy clusters are compared to observations of the Virgo cluster. The difficulty rises when observational probes depend on the formation history of the observed cluster. The diversity of clusters makes the one-to-one comparison with numerical simulations a daunting task to disentangle nurture versus nature. To remove the nurture from the study and focus only on the nature, it is necessary to produce a numerical cluster within the proper environment. Only then the tool is optimal to study the great physics laboratory that is the Virgo cluster.\\

The goal of this first paper of a series is to present the first full zoom-in hydrodynamical simulation of a counterpart of the Virgo cluster of galaxies, a \clone -- Constrained LOcal \& Nesting Environment -- simulation of Virgo, that has been obtained by constraining initial conditions with only local galaxy radial peculiar velocities. The large scale environment of the cluster resembles our local neighborhood. The zoom-in region has a $\sim$15~Mpc radius with an effective number of 8192$^3$ particles for the highest level (particle mass of  $\sim$3$\times$10$^7$~M$_\odot$ ) and a refinement down to 0.35~kpc. AGN and supernova feedbacks are included. This paper confirms that overall the simulated cluster reproduces the observed cluster well enough for future deeper comparisons of galaxy population and hot gas phase in the inner core.\\

First, general comparisons between the observed galaxy population of the Virgo cluster and the simulated one reveal excellent agreements in terms of luminosity and mass distribution as a function of the cluster center. A focused comparison between M87 and its simulated counterpart confirms the similarities. M87 counterpart is perhaps a bit too massive but some of this mass might be due to residual extinction in the observation and most probably a slightly too weak feedback in the simulation.  \\

Observationally-driven formation and evolution scenarios are then confronted with the theory down to the details. Our study confirms the number of small galaxies that entered the cluster within the last few Gigayears and the last significant merger. About 300 small galaxies (M$^*$$>$$10^7$M$_\odot$) entered the cluster recently, mostly within the last 500~Myr. In our previous dark-matter-only studies, we showed that the simulated Virgo cluster accreted matter along a preferential direction, a filament, and that the last significant merger occurred during the last 3-4~Gyrs. The group that finished merging within the last Gigayear or so was about 10\% the mass of the cluster today. Recently, \citet{2018ApJ...865...40L} proposed an observationally-driven merging scenario according to which a group of about 10\% Virgo counterpart mass entered the cluster along the line-of-sight preventing us from seeing it directly today. They found it identifying in a phase-space diagram a clump of red galaxies with high velocities relative to the cluster center but not in the inner core of the cluster. We identify the same group of galaxies in our simulation confirming both the result obtained in the dark-matter-only regime and the observational findings. The group entered the cluster via the filament diametrically opposed to the center of the box where the Milky Way counterpart is. The phase-space diagram produced with the simulated galaxies reveals an identical clump when positioning an observer in the line-of-sight of the infall direction of the group. \\

Overall, findings regarding the Virgo cluster, observational and numerical matches and residual discrepancies between the observed cluster and its counterpart can be summarized as follows:
\begin{enumerate}
\item Galaxies in the cluster core are in general redder and older.
\item The simulated Virgo cluster has an overall dynamical galaxy distributions matching the observed one with a virial radius of about 2.2 Mpc and a zero velocity radius of about 4.4 Mpc.
\item The galaxy luminosity and stellar mass distributions in the simulated Virgo cluster are in quite good agreement with the observed ones. The only exception being the inner core and the edges. However, edges might be affected by distance estimate uncertainties in the observational sample and the brightness of M87 might impact the observed inner core, observers deciding against observing them upon selection.
\item NUV-luminosity and stellar mass functions within respectively 0.15 Mpc and $\sim$3~Mpc obtained with the simulation match remarkably well the observations but at the faintest end in the first case. The mass resolution limit in the second case seems to be the only point of departure from the observational relation in the second case.
\item The counterpart of M87 is quite a good match to the observed one. Still small discrepancies are visible and cannot be entirely accounted for by aperture radii: M87 clone appears slightly too massive (bright) and not enough red with respect to the real M87. Its stars are also too metal rich although of quite the proper age. Explorations will be made later regarding for instance the growth of the black hole in M87 clone. Although more than big enough (M$>$10$^{11}$M$_\odot$), the black hole indeed grew recently because of a merger. However, since a detailed comparison between the observed and the simulated black holes will have to be conducted carefully, in particular because of the physics used for the AGN feedback, for instance the cosmic rays physics is not included, it will be the subject of a future study.
\item There are two good candidates for being M49 and M60 clones though they also are slightly too blue.
\item The Virgo cluster counterpart has had a quiet merging history within the last few Gigayears while it used to grow faster.
\item More precisely, Virgo counterpart underwent too major depressions in terms of galaxy accretions: first, about 7 Gyrs ago, when it transitioned from an active agglomerator to a quieter one in agreement with its cluster-size and close by environment; second, about 3 Gyrs ago, when a small group congested the filament located directly on the opposite side from ours, thus preventing it from channeling matter onto the cluster.
\item Virgo counterpart accretes matter along that filament, a preferential direction.
\item The last significant merging event was with the small congesting group. This group finished entering a bit more than a Gigayear ago. It was about 10\% the mass of the cluster today.
\end{enumerate}

This excellent numerical replica of the Virgo cluster of galaxies will permit studying more thoroughly the galaxy population of the cluster as well as the different galaxy types (jellyfish, backsplash, the fall-in small group, M87, etc) in the cluster and comparing them directly with their observed counterparts to test different hydrodynamical modelings. Future studies will address the hot gas phase of the simulated cluster. Virgo is known to be a cool-core cluster and this property is most probably due to its history. It will be an excellent way of testing baryonic physics in the intracluster medium. Several properties like metallicity, temperature obtained in our simulated cluster are precious probes to investigate. The simulation of the Virgo cluster opens great prospectives and let us foresee multiple interesting projects. Moreover, a companion simulation to tackle residual differences highlighted in this paper is already currently running using the same kinetic stellar feedback but with metal yield curves and mass return times based on a Kroupa IMF.

\section*{Acknowledgements}
This research has made use of the SIMBAD database, operated at CDS, Strasbourg, France as well as of the Extragalactic Distance Database (http://edd.ifa.hawaii.edu). The authors gratefully acknowledge the Gauss Centre for Supercomputing e.V. (www.gauss-centre.eu) for providing computing time on the GCS Supercomputers SuperMUC at LRZ Munich. GY and AK acknowledge financial support from \textit{Ministerio de Ciencia, Innovaci\'on y Universidades / Fondo Europeo de DEsarrollo Regional}, under research grant PGC2018-094975-C21. AK further thanks Fran\c coise Hardy for `tant de belles choses'.

\section*{Data availability}
The authors will gladly share data regarding the galaxy population and dark matter halos of the \clone\ simulation of the Virgo cluster upon request.


\bibliographystyle{mnras}

\bibliography{biblicompletenew}
 \label{lastpage}

\section*{Appendix}
\renewcommand{\thesubsection}{\Alph{subsection}}
\subsection{DM vs. Hydro run}

\begin{figure}
\centering
\includegraphics[width=0.45 \textwidth]{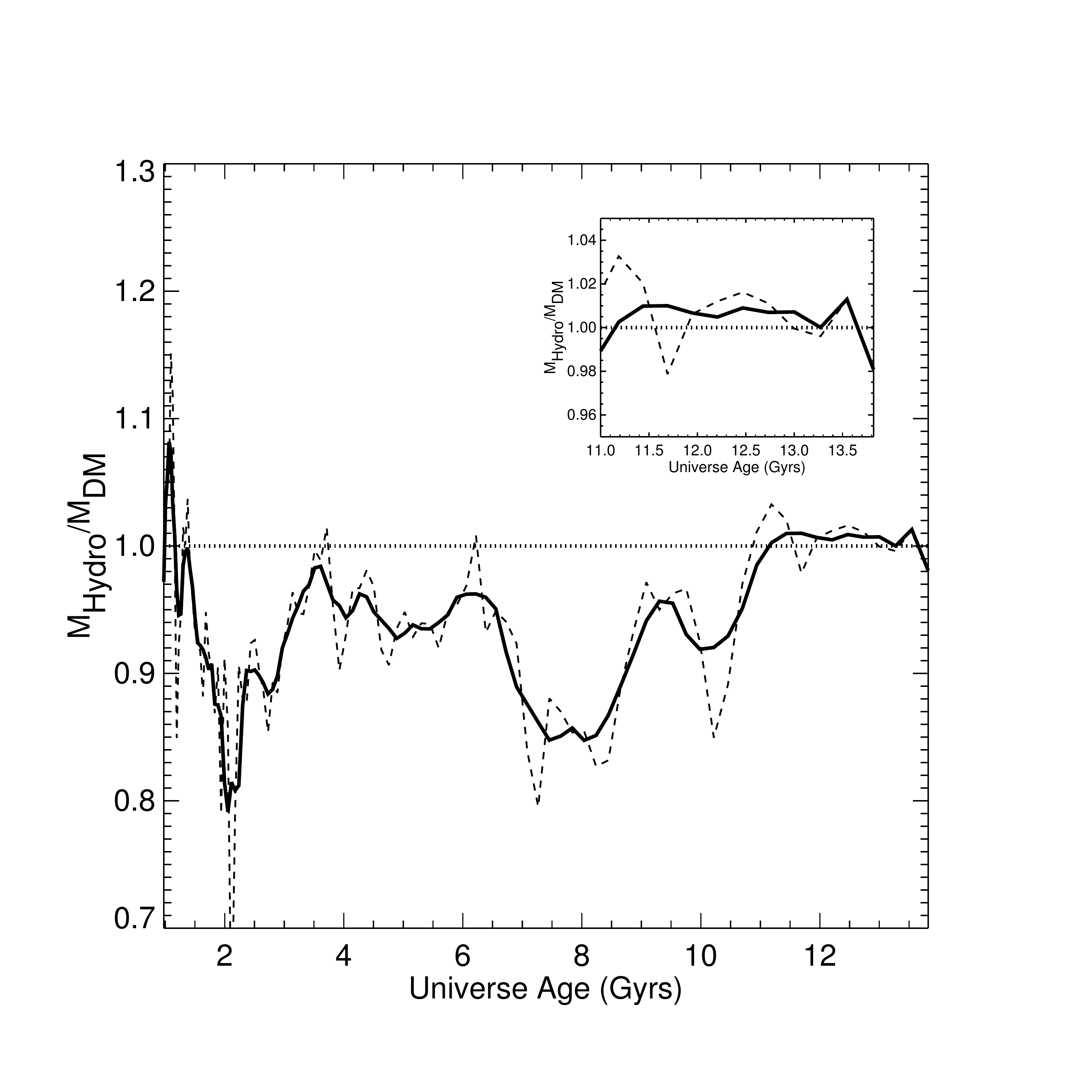} \\
\vspace{-1.5cm}

\includegraphics[width=0.45 \textwidth]{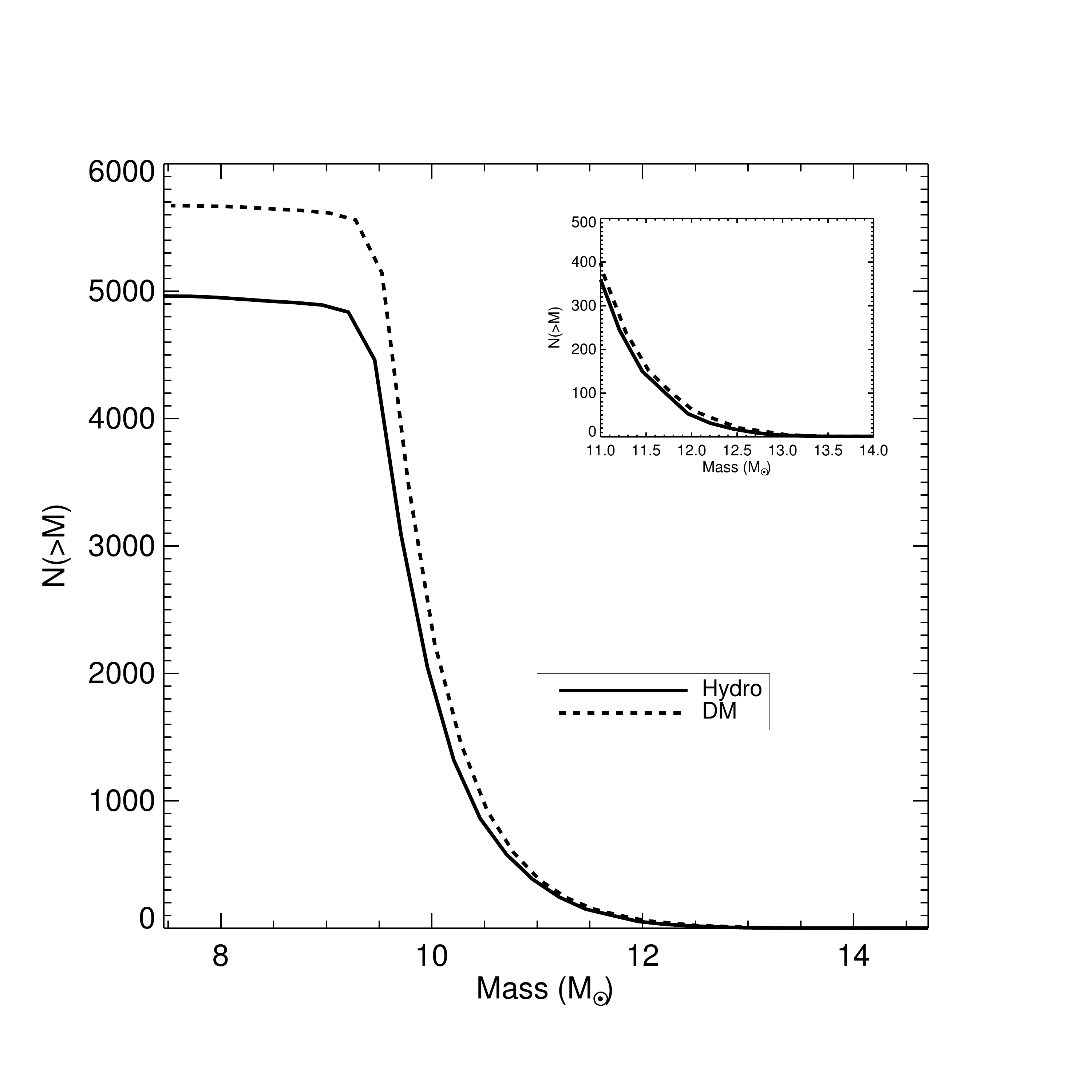}
\caption{Comparisons between two runs with the same parameters and same initial conditions but one with gas (Hydro), the other without (DM). Top: Ratio of the total virial mass of the halos in both runs as a function of time (after accounting for the baryonic content in the hydrodynamical run). The dashed line is obtained considering each snapshot while the solid line is a mean across five snapshots. Bottom: Halo mass functions in both runs. Adding gas suppresses small mass halos but does not impact the high mass end of the function as expected.}
\label{fig:dmhydro}
\end{figure}

This appendix intends to demonstrate that the hydrodynamic component of the simulation did not perturb beyond its normal effect the dark matter result. Figure \ref{fig:dmhydro} top confirms that the final virial mass of the cluster is similar with and without gas after rescaling to account for the baryonic content and that there are only moderate changes in the assembly history especially recently. The bottom of the same figure shows that indeed hydrodynamics suppresses the small mass halos but does not affect the high mass end of the dark matter halo mass function.

\subsection{Cluster-centric velocities}

\begin{figure}
\centering
\includegraphics[scale=0.35,angle=-90]{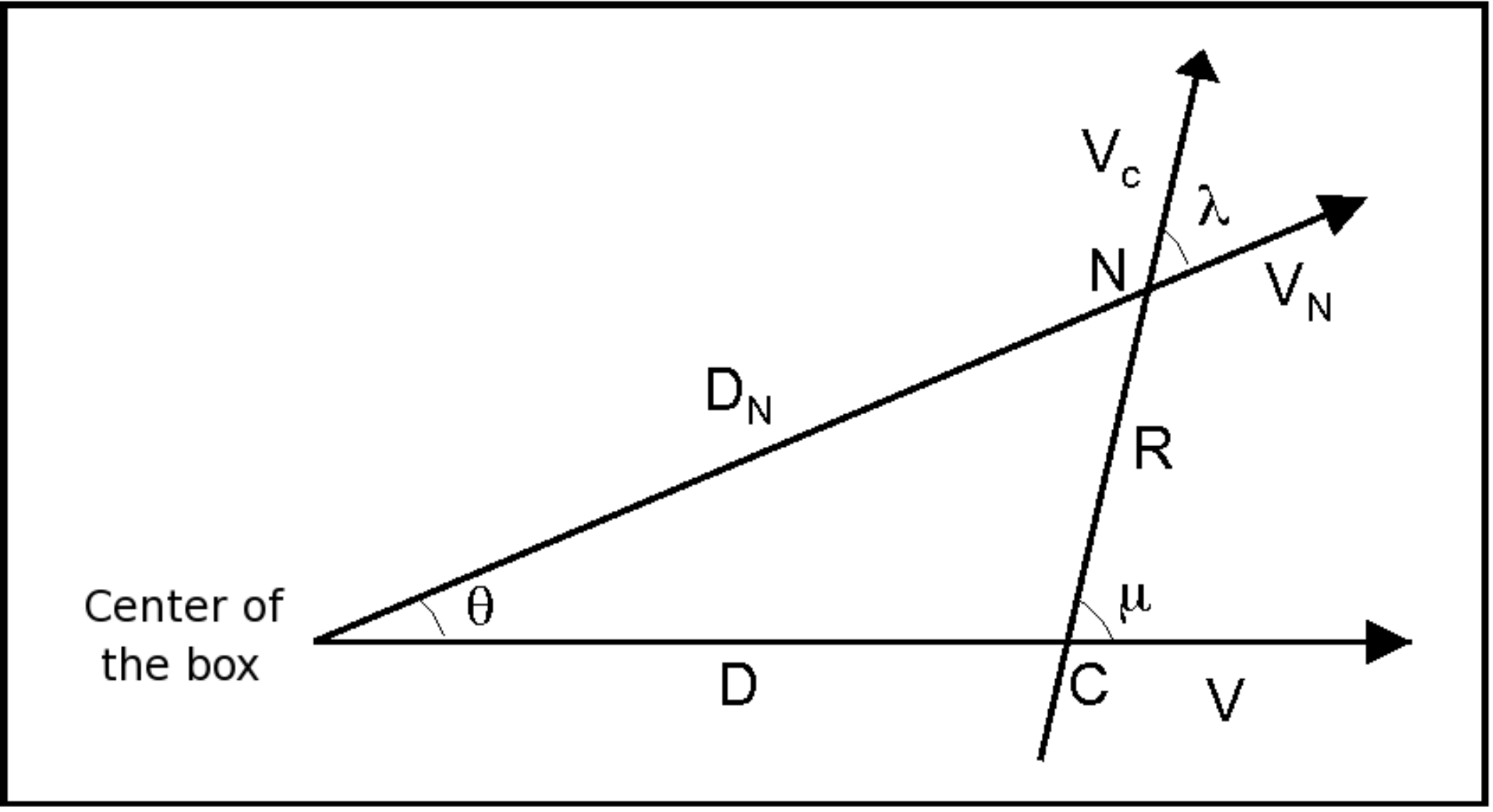}
\caption{Diagram representing the different parameters required to compute the distance and the velocity of a galaxy with respect to the center of a system.}
\label{schema}
\end{figure}

To compute cluster-centric velocity with the help of Figure \ref{schema} and \citet{2006Ap.....49....3K}, let C be the center of a system of galaxies located at the distance D and receding with a velocity V from us. Consider a galaxy N at the distance D$_N$ and receding at V$_N$ from us. We call the angular separation, between the center of the system and the galaxy N, $\theta$.  Then, the distance between the galaxy N and the center of the system R$_c$ can be expressed as follows:
\begin{equation}
R_c=R=\sqrt{D^2 + D_N^2 - 2 D D_N cos \theta}
\end{equation}
The galaxy N is going away from C at the velocity:
\begin{equation}
V_c=V_N cos \lambda - V cos \mu
\end{equation} 
where $\mu = \lambda + \theta$ and tan $\lambda$ = $\frac{D sin \theta}{D_N - D cos \theta}$ assuming that random peculiar velocities are low compared to expansion velocities.

\subsection{Zero velocity radius}
The zero velocity radius is a priori the intersection of the running median with the X-axis. However, because of a high velocity dispersion in cluster cores, running medians are noisier close to the cores than in the far outskirts where the Hubble flow is reached. Consequently, we fit the running median with a theoretical velocity profile derived from the spherical model by \citet{2008A&A...488..845P}. This profile assumes a spherical collapse model, $\Lambda$ included, in the outskirts of the clusters. In the core, assumed to contain most of the mass where shell crossing has already happened, the orbits are mainly radial. Then:
\begin{equation}
V(R_c)=1.377 H_0 \times R_c - 0.976 \times \frac{H_0}{R_c^n} \times \left (\frac{GM}{H_0^2}\right )^{\frac{n+1}{3}}
\end{equation} 
with M the core cluster mass, R$_c$ the distance of the galaxy to the cluster center, V(R$_c$)=V$_c$ the radial velocity of the galaxy with respect to the cluster center and n=0.627 at z=0. By definition, the velocity at the zero velocity radius is null. Because the model is valid only up to the point where the Hubble Flow is reached, fits are based on the left end of the running median where cores and outskirts are in close contact with each other. We proceed iteratively until the portion of the running median to fit is delimited such that the model converges to the data.\\

\subsection{Completeness of the simulation}

\begin{figure}
\hspace{-0.4cm}\includegraphics[width=0.48 \textwidth]{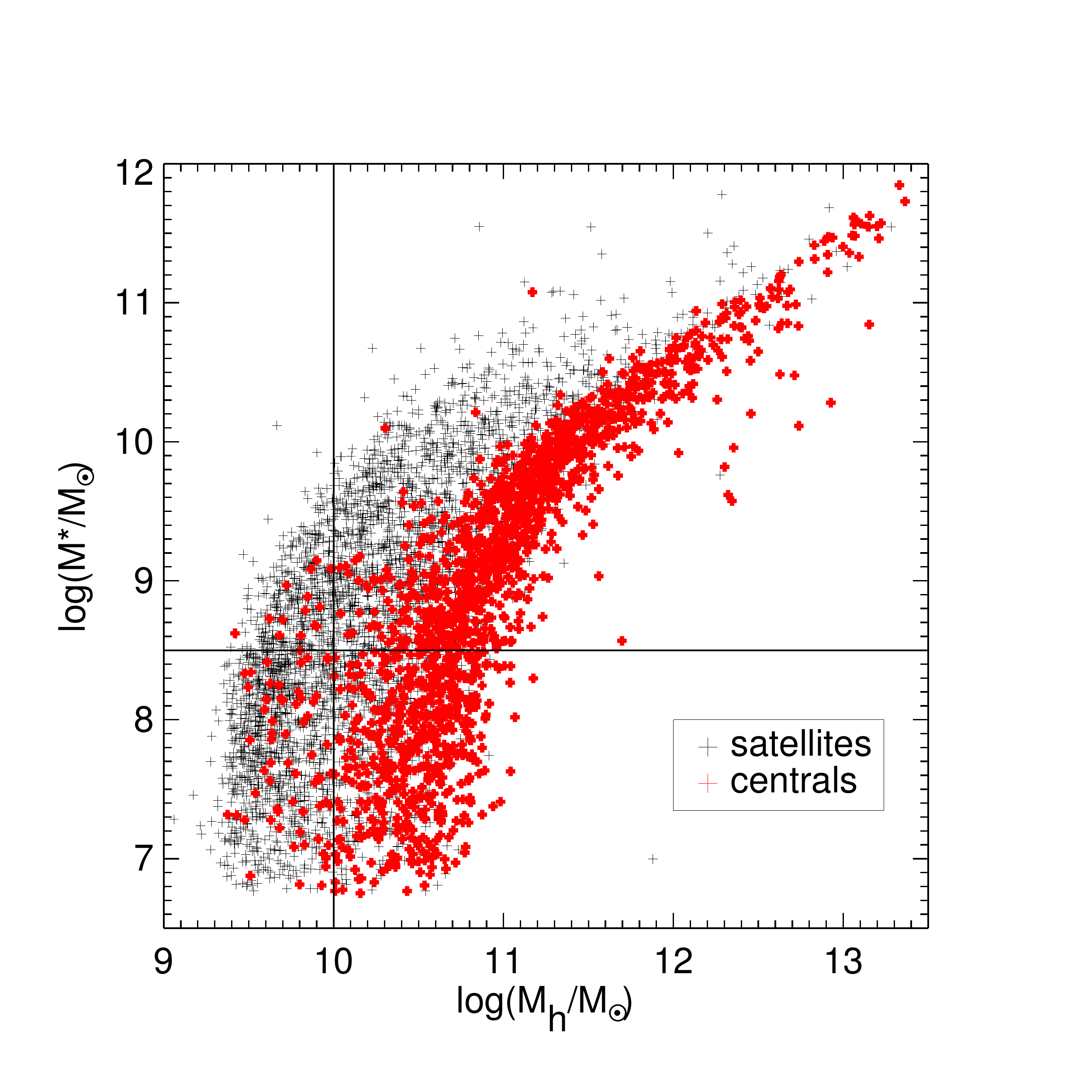}
\vspace{-1.6cm}

\hspace{-0.3cm}\includegraphics[width=0.47 \textwidth]{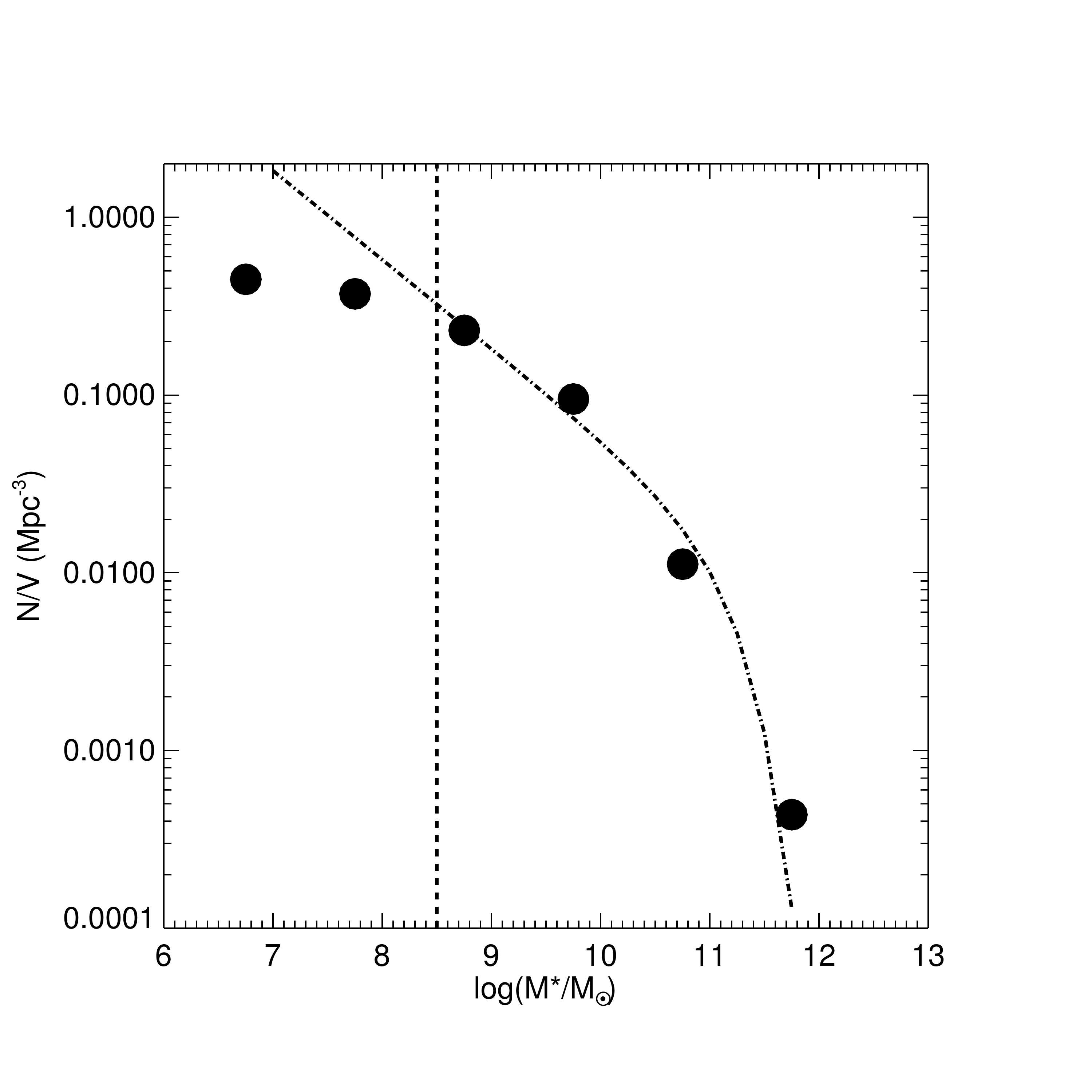}
\vspace{-0.2cm}
\caption{Top: Stellar mass as a function of dark matter halo mass for central (red crosses) and satellite (black crosses) galaxies in the 12~Mpc radius spherical volume. Bottom: Galaxy stellar mass function of the 12~Mpc radius spherical volume (filled circles) fitted with a double Schechter function (dot-dashed line). The completeness is visible at $\mathrm{M}^*=10^{8.5}$M$_\odot$.}
\label{fig:massres}
\end{figure}

Figure \ref{fig:massres} top shows the completeness of the simulation at $\mathrm{M}^*=10^{8.5}$M$_\odot$. This sets the limit of validity of the close comparisons between observations and simulations in this paper without correction for completeness. To evaluate the correction, we fit a double Schechter function to the galaxy stellar mass function within the 12~Mpc radius spherical volume of the simulation using only points above the completeness limit \citep[e.g.][]{2020arXiv200903176M}:
\begin{equation}
\Phi(M)=\mathrm{ln}(10) \mathrm{exp}[-10^{(M-M^*)}]10^{(M-M^*)}[\Phi^*_1 10^{(M-M^*)\alpha_1}+\Phi^*_2 10^{(M-M^*)\alpha_2}]
\end{equation}
 
We use as a first proxy the parameters from \citet{2012MNRAS.421..621B} who derived galaxy stellar mass functions in the local Universe. We increase $M^*$ following \citet{2013MNRAS.432.3141C} who found that $M^*$ evolves with the environment density and $\Phi^*_2$ following \citet{2008MNRAS.388..945B} who observed steeper slopes at the low mass end in clusters than in the field. These parameters are adjusted to fit only the points above the completeness limit as shown in Figure \ref{fig:massres} bottom. Finally,  $M^*$=11, log$(\Phi^*_1)$=-2.4, log$(\Phi^*_2)$=-2.1, $\alpha_1$=-0.35, $\alpha_2$=-1.5.\\

Typically about 10,000-12,000 (50-60) small galaxies of $\mathrm{M}^*~>~10^{7}$M$_\odot$ are missed in the 12~Mpc radius spherical volume (simulated Virgo cluster) given the parameters used for the halo finder run and the resolution of the simulation.

\end{document}